%% file: main.tex
\documentclass[prx,twocolumn,aps,superscriptaddress,nofootinbib,floatfix]{revtex4-2}

\usepackage[T1]{fontenc}
\usepackage[utf8]{inputenc}
\usepackage[british]{babel}
\IfFileExists{lmodern.sty}{\usepackage{lmodern}\usepackage{microtype}}{}
\usepackage[margin=2.5cm]{geometry}
\usepackage{graphicx}
\usepackage{booktabs}
\usepackage{caption}
\usepackage{amsmath}
\usepackage{xurl}
\usepackage{orcidlink}   
\hypersetup{hidelinks}

\usepackage{url}
\usepackage{lineno}

\input{generated/stats_macros.tex}      
\input{generated/results_macros.tex}    
\input{generated/fragmentation_macros.tex}
\input{generated/methods_macros.tex}
\input{generated/effort_macros.tex}     
\input{generated/removal_macros.tex}    
\input{generated/table_macros.tex}      
\input{generated/derived_macros.tex}    
\input{generated/sources_macros.tex}    


\begin{document}

\title{What It Costs to Trust a Source: A Reproducible Method for Institutional Assessment of Author Disambiguation}

\author{Simon J. Porter\,\orcidlink{0000-0002-6151-8423}}
\email{s.porter@digital-science.com}
\homepage{\mbox{https://orcid.org/0000-0002-6151-8423}}
\affiliation{Digital Science, 6 Briset Street, London EC1M 5NR, United Kingdom}

\date{29 October 2026}

\begin{abstract}
\noindent
Records of research outputs are curated with care by universities, and the effort is shaped by the external sources from which candidate records are drawn. The curation required before the author identities in such sources can be trusted is examined, and a reproducible method for estimating its cost is set out. Four tests are described; two can be run on public data alone, and two are conducted against an institution's own curated record, taken from Symplectic Elements Discovery portals. The method is demonstrated on OpenAlex and Dimensions across \statNPrimary{} institutions. Of the publications attributed to matched researchers, \statOAConfirmedPctPrimary\% were confirmed by the institutional record for OpenAlex and \statDimConfirmedPctPrimary\% for Dimensions, and fewer trusted researcher profiles were retained from OpenAlex at a matched confidence threshold in each of the \statNPrimary{} institutions. Carried through to effort, integrating OpenAlex requires \effBacklogRatioHalf{} times the person-days of review and entry of Dimensions even when each feed is filtered at the threshold that minimises the amount of review required.
\end{abstract}

\maketitle


\input{sections/01_introduction}
\input{sections/02_related_work}
\input{sections/03_methods}
\input{sections/04_results}
\input{sections/05_discussion}
\input{sections/06_limitations}
\input{sections/07_conclusion}
\input{sections/08_endmatter}

\bibliographystyle{plainnat}
\bibliography{references}

\end{document}

%% file: generated/stats_macros.tex
\newcommand{\statOAOverPct}{39.2}
\newcommand{\statOAUnderPct}{23.3}

\newcommand{\statOAMeanDev}{3.99}
\newcommand{\statOANAuthors}{176{,}699}
\newcommand{\statDimOverPct}{12.9}
\newcommand{\statDimUnderPct}{9.9}
\newcommand{\statDimEqualPct}{77.2}
\newcommand{\statDimMeanDev}{0.24}
\newcommand{\statDimNAuthors}{80{,}256}
\newcommand{\statOATailN}{137{,}947}
\newcommand{\statDimTailN}{4{,}166}
\newcommand{\statNAll}{51}
\newcommand{\statNPrimary}{30}
\newcommand{\statOAConfirmedPctAll}{50.6}
\newcommand{\statOAConfirmedPctPrimary}{50.4}
\newcommand{\statDimConfirmedPctAll}{69.8}
\newcommand{\statDimConfirmedPctPrimary}{69.6}

\newcommand{\statOANotInSystemPctPrimary}{46.6}

\newcommand{\statDimNotInSystemPctPrimary}{26.7}
\newcommand{\statRatioBelowOneAll}{51}

\newcommand{\statStdPrimary}{0.027}
\newcommand{\statRatioMinPrimary}{0.85}
\newcommand{\statRatioMaxPrimary}{0.96}

\newcommand{\statMedianRatioPrimary}{0.889}

%% file: generated/results_macros.tex
\newcommand{\statOARecallPermissive}{70}
\newcommand{\statDimRecallPermissive}{86}
\newcommand{\statOAFirstStepPct}{9}
\newcommand{\statDimFirstStepPct}{5}
\newcommand{\statRecallDisjointFrom}{0.0}
\newcommand{\statRetentionDisjointFrom}{0.2}
\newcommand{\statOAExpRemovedPct}{58}
\newcommand{\statOAConfLostPct}{6}
\newcommand{\statDimExpRemovedPct}{50}
\newcommand{\statDimConfLostPct}{4}
\newcommand{\statOASinglePubPct}{56.3}
\newcommand{\statDimSinglePubPct}{16.7}

%% file: generated/fragmentation_macros.tex
\newcommand{\statOAMultiIdPct}{35.1}
\newcommand{\statDimMultiIdPct}{19.6}

\newcommand{\statOAMultiIdPctFiltered}{32.8}
\newcommand{\statDimMultiIdPctFiltered}{18.7}

%% file: generated/methods_macros.tex
\newcommand{\countryAllUS}{15}

\newcommand{\countryAllGB}{13}

\newcommand{\countryAllAU}{8}

\newcommand{\countryAllNZ}{3}

\newcommand{\countryAllIE}{2}

\newcommand{\countryAngloAll}{44}
\newcommand{\countryAngloPrimary}{28}
\newcommand{\fragOAOrcids}{168{,}960}
\newcommand{\fragOAMultiOrcids}{6{,}795}
\newcommand{\fragOAMultiOrcidsPct}{4.0}
\newcommand{\fragOARowsMultiPct}{8.2}
\newcommand{\fragDimOrcids}{80{,}256}
\newcommand{\fragDimMultiOrcids}{0}

\newcommand{\sensUsersFortyPooled}{59{,}553}
\newcommand{\sensUsersTwentyPooled}{56{,}752}
\newcommand{\sensCeilingDropPct}{4.7}

%% file: generated/effort_macros.tex
\newcommand{\effVerifySecLow}{10}
\newcommand{\effVerifySec}{15}
\newcommand{\effVerifySecHigh}{20}
\newcommand{\effCheckSec}{35}
\newcommand{\effCheckSecHigh}{74}
\newcommand{\effRejectMinLow}{1}
\newcommand{\effRejectMin}{5}
\newcommand{\effRejectMinHigh}{15}
\newcommand{\effEntryMin}{3}
\newcommand{\effEntryMinHigh}{10}
\newcommand{\effResDecideSec}{15}
\newcommand{\effOverheadMin}{2}
\newcommand{\effOverheadMax}{10}
\newcommand{\effMinutesPerDay}{435}

\newcommand{\effWindowStart}{1 October 2025}
\newcommand{\effWindowEnd}{30 September 2026}
\newcommand{\effTrueRate}{296{,}857}
\newcommand{\effOADimClockPct}{99.7}
\newcommand{\effOAUnconfLatePct}{2}
\newcommand{\effDimUnconfLatePct}{6}
\newcommand{\effOABacklogNone}{47{,}486}
\newcommand{\effOABacklogHalf}{15{,}410}

\newcommand{\effOABacklogHalfLow}{13{,}815}

\newcommand{\effOAResFlowNone}{2{,}482}
\newcommand{\effOAResFlowHalf}{2{,}348}
\newcommand{\effOAOptThreshold}{0.5}
\newcommand{\effOAFlatLow}{0.3}
\newcommand{\effOAFlatHigh}{0.6}
\newcommand{\effOAFilterSavingPct}{68}

\newcommand{\effOASuggestionsPerYear}{4.4}
\newcommand{\effOASessionsWeekly}{4.0}

\newcommand{\effOAItemsPerSession}{1}
\newcommand{\effOADigestSavingPct}{13}
\newcommand{\effOAUnconfWeakIdPct}{69}
\newcommand{\effOACeilTwentyRisePct}{30}

\newcommand{\effOAHighFactor}{3}
\newcommand{\effDimBacklogNone}{19{,}878}
\newcommand{\effDimBacklogHalf}{10{,}159}

\newcommand{\effDimBacklogHalfLow}{8{,}545}

\newcommand{\effDimResFlowNone}{1{,}839}
\newcommand{\effDimResFlowHalf}{1{,}756}
\newcommand{\effDimOptThreshold}{0.5}
\newcommand{\effDimFlatLow}{0.3}
\newcommand{\effDimFlatHigh}{0.6}
\newcommand{\effDimFilterSavingPct}{49}

\newcommand{\effDimSuggestionsPerYear}{4.4}
\newcommand{\effDimSessionsWeekly}{3.6}

\newcommand{\effDimUnconfWeakIdPct}{35}
\newcommand{\effDimCeilTwentyRisePct}{24}

\newcommand{\effDimHighFactor}{3}
\newcommand{\effBacklogRatioHalf}{1.5}
\newcommand{\effBacklogRatioNone}{2.4}
\newcommand{\effFlowRatioHalf}{1.8}
\newcommand{\effResFlowRatioHalf}{1.34}

\newcommand{\effResFlowRatioHalfZero}{1.98}
\newcommand{\effResFlowRatioHalfMax}{1.07}
\newcommand{\effResFlowGapHalf}{592}
\newcommand{\effResFlowGapHalfZero}{625}
\newcommand{\effResFlowGapHalfMax}{462}
\newcommand{\effOptThresholdInvestLow}{0.2}
\newcommand{\effOptThresholdInvestHigh}{0.7}
\newcommand{\effOAKeepsDimNewPct}{79}
\newcommand{\effDimKeepsOANewPct}{96}
\newcommand{\effNewWorksDimN}{189{,}718}
\newcommand{\effNewWorksOAN}{112{,}830}
\newcommand{\effOAMissNewIdPct}{66}
\newcommand{\effOAMissAbsentPct}{8}
\newcommand{\effOAMissNewIdSmallPct}{93}
\newcommand{\effResGapFastEntryHighLogin}{462}
\newcommand{\effResRatioFastEntryHighLogin}{1.07}
\newcommand{\effResGapSlowEntryNoLogin}{2{,}227}
\newcommand{\effResRatioSlowEntryNoLogin}{2.24}
\newcommand{\effResGapSlowEntryHighLogin}{2{,}065}

%% file: generated/removal_macros.tex
\newcommand{\statOASelfCollabN}{80{,}951}
\newcommand{\statOASelfCollabRemoved}{1{,}703}
\newcommand{\statOASelfCollabPct}{2.1}
\newcommand{\statOASelfCollabWorksPer}{647}
\newcommand{\statOAKeptWorksPer}{91}
\newcommand{\statOASelfCollabWorksSharePct}{13}
\newcommand{\statOASelfCollabConfPct}{36.4}
\newcommand{\statOAKeptConfPct}{50.4}
\newcommand{\statOAWithRemovedConfPct}{48.6}
\newcommand{\statDimSelfCollabN}{80{,}809}
\newcommand{\statDimSelfCollabRemoved}{1{,}063}
\newcommand{\statDimSelfCollabPct}{1.3}
\newcommand{\statDimSelfCollabWorksPer}{442}
\newcommand{\statDimKeptWorksPer}{73}
\newcommand{\statDimSelfCollabWorksSharePct}{8}
\newcommand{\statDimSelfCollabConfPct}{54.5}
\newcommand{\statDimKeptConfPct}{69.6}
\newcommand{\statDimWithRemovedConfPct}{68.5}

%% file: generated/table_macros.tex
\newcommand{\repBaselinePrimary}{4{,}340{,}379}
\newcommand{\repBaselineFull}{4{,}597{,}719}

%% file: generated/derived_macros.tex
\newcommand{\statTailRatio}{33}
\newcommand{\statExposureRatio}{1.7}
\newcommand{\statDimMoreProfilesPct}{12}
\newcommand{\statOverRatio}{3}
\newcommand{\effBacklogGapHalfLow}{5{,}300}

%% file: generated/sources_macros.tex
\newcommand{\srcAnalysisDate}{25 October 2026}
\newcommand{\srcOpenAlexRelease}{14 October 2026}
\newcommand{\srcDimensionsRelease}{25 October 2026}
\newcommand{\srcCrossrefRelease}{7 October 2026}
\newcommand{\srcDiscoveryHarvest}{26 May 2026}
\newcommand{\srcOrcidLoaded}{21 October 2026}

%% file: sections/01_introduction.tex
\section{Introduction}\label{sec:intro}

Since the early 2000s, bibliographic sources on the web have helped universities to assemble records of the research of their staff. In most research intensive institutions Current Research Information Systems (CRIS) now support compliance and assessment submissions, open-access monitoring, annual reporting and the public profiles through which an institution presents its people to collaborators, funders and the press \citep{bryant2018}. The outcomes based on these uses are consequential for both individuals and institutions and hence records are curated in collaboration and with care by research offices, librarians and researchers.  This activity is both ongoing and significant and hence the ongoing cost of curation is a recurring concern in the literature on research information management \citep{azeroual2019}.

The scholarly records upon which this curation is based have been built in layers: Crossref has registered DOIs and publication metadata that has been deposited by source publishers since 2000 \citep{hendricks2020} and has made efforts to include historic content; and DataCite, founded in 2009 as a separate DOI registration agency, brought datasets, and later other research outputs, into the DOI system \citep{brase2009}. ORCID launched in 2012 to give research participants a public record of their outputs that they maintain themselves \citep{haak2012}. In each of these registries an assertion is made by a party with direct knowledge of a fact concerning a scholarly record, and each claim carries clear provenance, in the sense of FAIR principle R1.2 \citep{wilkinson2016}. To be explicit, we interpret this to mean not only that every metadata item should conform to the FAIR principles, but that within reason, the provenance of the assertion should be implicitly clear. That is to say that, for example, metadata vested on Crossref can be assumed to be findable by virtue of Crossref's data services; accessible through Crossref's use of standardised, open data formats; interoperable as a result of the aforementioned formats; and reusable not only due to Crossref's licence terms but also implicitly because the record is trustworthy, being as provided to Crossref by its originator. This trust in: i) faithful reproduction of the record from a source; ii) the ability to assume such provenance for a whole record without the need to check individually; and iii) the knowledge that records have not been added or deleted and that the data provided by Crossref constitutes an unfiltered and unredacted set is critical to all the layers that then derive from it.

A second layer of sources, among them Web of Science, Scopus and Dimensions \citep{visser2021} and, more recently, OpenAlex \citep{priem2022}, constructs knowledge that the registries do not hold: algorithmic identifiers for researchers whose names appear only as text, affiliations resolved from address strings, and classifications by field.

In this context a \emph{secondary source} is an index that derives assertions about authorship algorithmically from metadata that it did not itself author, whether or not those assertions are later curated by hand. By a \emph{person-level record} we mean the attribution of research outputs to a specific individual, expressed as an author identifier together with the outputs gathered under it, as distinct from the document-level record of what was published, where and with what citations, on which most comparisons of sources have concentrated. There are two kinds of failure that are common but which are not equally visible. A missing research output in a person's record is a gap that someone eventually notices, whereas a researcher profile that contains another person's work looks to every downstream system like a richer profile, even if the researcher appears to be publishing in a different domain with different collaborators, this is not unusual - many researchers change fields during their career---it might also mark a transistion to a different field of research. These ``richer'' profiles have an increasingly consequential impact due to the growing class of applications, from research-intelligence platforms that match researchers to policy-engagement opportunities \citep{overton2026} to integrity screens that flag anomalous publication behaviour \citep{ioannidis2018,porter2024networks}, and all these applications broadly take author identities as given and author disambiguation if not as a solved problem then a sufficiently solved problem.

An institution that brings such a source into its CRIS therefore takes on work that is not usually included in the calculation of the cost of ownership. We call it the \emph{integration cost}, and measure it in effort: the curation that must be layered on top of a source before its claims about the institution's own people can be acted upon. Both the level of curation and where that curation lands is a matter of choice - informed by institutional and national cultural norms, financial considerations, trust in systems, trust in data, local ethical and regulatory practices, risk tolerance, and intended data use. Effort spent correcting a secondary source repairs one copy of the record inside another organisation's database, whereas confirmed person--publication links maintained in the institution's own CRIS fix the institutional record and, where connected through ORCID, supply an authority from which every downstream index can draw.

This paper sets out a reproducible method for estimating the integration cost of a scientometric source from data that an adopting institution already holds. Two of its four tests are \emph{source-level}, using only a source's output and the ORCID public data file to assess how likely the source is to conflate distinct researchers; two are \emph{institutional}, reading the source against an institution's own curated record as a proxy for a golden dataset. The method is demonstrated on OpenAlex \citep{priem2022} and Dimensions \citep{herzog2020,hook2018}, two indexes built on open persistent-identifier infrastructure and openly available metadata, across \statNPrimary{} institutions whose records are exposed through Symplectic Elements Discovery portals. The results are expressed in person-days, for administrators reviewing a feed and for researchers reviewing their own suggestions.

\begin{figure}[t]
\centering
\includegraphics[width=\columnwidth]{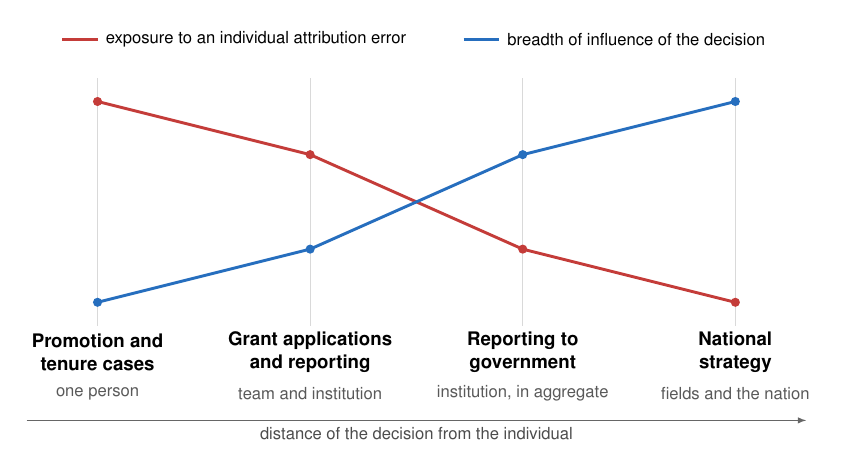}
\caption{Uses of the institutional publication record, ordered by the distance of the decision from the individual researcher. The schematic is qualitative: the positions are ordinal and the crossing point is illustrative rather than measured.}
\label{fig:uses}
\end{figure}

The publication data held by a university are put to uses that differ considerably in how much error they can tolerate (Figure~\ref{fig:uses}). At one extreme sits the case for promotion. In this case records are read by a committee whose decision bears directly on an individual's career; a misattributed or missing paper may materially affect the outcome and is a matter on which the individual might reasonably seek redress. The grant application is not far removed. A funder will expect an accurate account of the outcomes of earlier awards and of the work of each investigator named in the proposal, and an error reflects not only on the individuals concerned but on the standing of the institution with that funder.

Reporting to government occupies a middle position. Accuracy is still required, but most evaluative systems treat the data in aggregate, where an individual misattribution is diluted within the totals of a department or an institution; an exercise such as the UK's Research Excellence Framework, in which named outputs are submitted against named staff, sits rather closer to the stricter end than its aggregate presentation might suggest. At national level the data are used to discern trends: whether a field is growing, whether investment should follow, and how a research system aligns with a government's view of critical technologies. These are among the most consequential allocations made from such data, and yet the least sensitive to the accuracy of any individual record.

It follows that the bodies with the greatest influence over the direction of research funding are, in some sense, the least exposed to the consequences of inaccurate attribution. A national decision stands one step removed from the institutions that carry its consequences and two steps removed from the individuals whose records underlie it, while the cost of curation, and the risk that attends its absence, fall on research offices and researchers.

Aggregation is forgiving of random error, though rather less so of systematic error, and profile conflation is of the latter kind. It concentrates on common names, which may be expected to cluster by naming convention and hence by country and region, and possibly by field, which are the units that national analyses compare, so whether a source's errors lean in one direction also matters beyond the research office. For the institution itself, the confidence required of a source becomes a property of the use to which the data are put, and a single institution will in practice need several levels of confidence from the same feed.

This paper is organised as follows: We provide a brief literature review (Section~\ref{sec:related}) before describing the data and test methodology  in Sec.~\ref{sec:methods}. Sec.~\ref{sec:results} contains our results, after which we discuss their implications in Sec.~\ref{sec:discussion}) and note the limitations of our study in Sec.~\ref{sec:limitations}.

%% file: sections/02_related_work.tex
\section{Related work}\label{sec:related}

\textbf{The cost of research information.}  The uses of a CRIS have expanded steadily, from compliance and assessment through open-access monitoring and research-data registration to public researcher profiles \citep{bryant2018,schopfel2017}. Increasingly the CRIS also underpins what might be called \emph{university-in-the-loop} reporting: measurement grounded in records that the institution itself verifies and warrants, in contrast to rankings and evaluations computed entirely from secondary sources outside its control. The Barcelona Declaration on Open Research Information reinforces this direction, committing its signatories to make openness the default for the research information they use, including to assess researchers, and to require that the data in research information systems can be exported and made open, so that the information used in evaluation is accessible and auditable by those being assessed \citep{barcelona2024}, commitments that apply directly to institutional records such as those studied here.

A registry that drives public profiles and institutionally warranted reporting must be correct at the level of the individual researcher, and the literature is consistent that such correctness is bought with labour: profiling and cleansing routines \citep{azeroual2018}, quality problems reported across national contexts \citep{azeroual2019}, a standard of fitness for purpose that rises as uses accumulate \citep{gurney2022}, and work on integrating repositories and workflows \citep{scherer2020}. What has been missing is a measurement connecting this literature to the choice of external source: how far the curation burden of a CRIS is determined in advance by the disambiguation quality of the source that feeds it, and whether that burden differs between sources consistently enough to be weighed in the choice.

\textbf{The effort of record entry and verification.} No published study, to our knowledge, measures how long it takes to accept or reject a publication that a CRIS suggests, so the effort of integration has to be assembled from adjacent tasks. For entering a record by hand, those who run a CRIS work with figures of a few minutes: the Portuguese national CV platform assumes three minutes per manually entered record in estimating the effort saved by its suggestion service \citep{fccn2024cienciavitae}, a case study of open-access workflows in Symplectic Elements estimates three minutes to create a record with its full text \citep{symplectic2016qmul}, and repository guidance allows five to ten minutes for a complete deposit \citep{ua_ir_deposit}. 

Confirming an existing record has also been observed to be quicker than creating one. In a time-driven study of cataloguing, adapting an existing bibliographic description took 74 seconds against 306 seconds to create one, about three-quarters less, although across the whole workflow, physical processing included, the saving was a third \citep{siguenzaguzman2014cataloguing}. Screening for systematic reviews supplies the nearest logged measurements of a quick bibliographic judgement: a mean of 35 seconds per abstract over nearly 400,000 decisions on one platform \citep{howard2020swift}, and medians of 15 to 20 seconds in a small experiment reported by a vendor \citep{oblenis2017vote}. These are judgements of relevance against inclusion criteria rather than of authorship, and the full-text assessment that a doubtful case can require has been put, from one team's experience, at about fifteen minutes \citep{trad2025streamlining}. 

Administrative records may be maintained by researchers as well as administrators. A researcher who reviews suggestions personally also bears the overhead of logging into each session. Applying published second-factor times of roughly 10 to 21 seconds after the password to logs of more than 35 million two-factor logins at two universities, \citet{reynolds2020systemic} find more than one login attempt in twenty abandoned or failed at the second step, and locate the burden less in the total time than in the position of the step on the critical path to the task the user came to do.
 Administration can also be considered to be an interruption to researchers' other responsibilities. Without supplying a cost per session, research on interruption finds that interrupted work is often not resumed for tens of minutes, during which other work intervenes \citep{mark2005notask,iqbal2007disruption}, and people compensate for interruption by working faster while reporting greater stress \citep{mark2008cost}. Estimating the effort that research-information workflows consume, and the share of it that persistent identifiers can remove, has also been developed at national scale \citep{brown2022}.

\textbf{Comparing scientometric sources.}  OpenAlex \citep{priem2022} inherited both the corpus and the ambition of Microsoft Academic Graph \citep{sinha2015,wang2020}, a fully open, algorithmically constructed index of the scholarly record including machine-generated author identities. Dimensions took a related route, assembling a linked index around persistent identifiers with a disambiguation layer of its own \citep{hook2018,herzog2020}. The comparative literature across these and the established commercial indexes has concentrated largely on documents: \citet{visser2021} compared Scopus, Web of Science, Dimensions, Crossref and Microsoft Academic for document coverage and for the completeness and accuracy of citation links, and more recent work has examined OpenAlex's reference coverage \citep{culbert2025} and its coverage and metadata quality for African research \citep{alonso2025}. These studies establish with increasing precision what is \emph{in} each source, but say less about whether each source attributes that content to the right people.

The closest examination of that question for OpenAlex, a benchmark of its author identifiers against Clarivate's using a gold set assembled with the assistance of a large language model, is available at the time of writing only as a preprint \citep{zhao2025}. Evaluations of this kind share a construction in which a sample of author--publication assignments is labelled, by hand or by a stronger model, for the purpose of the evaluation. The ground truth used here differs in kind. Institutional records are maintained continuously for operational purposes, cover whole researcher populations rather than samples, and carry institutional consequences that discipline their accuracy. To our knowledge, the disambiguation of neither source has been evaluated against institutions' own records.

\textbf{Author name disambiguation and its failure modes.}  Author name disambiguation has a well-developed methods literature \citep{smalheiser2009,ferreira2012}, and its canonical failure taxonomy is the one adopted here: \emph{splitting}, in which one person's output is fragmented across several identities, and \emph{lumping}, in which several people are merged into one. Production systems manage the trade-off explicitly. \citet{caron2014} describe the rule-based scoring behind the CWTS system, tuned conservatively to prefer precision over recall; \citet{tekles2020} compare unsupervised approaches against a large gold standard; and \citet{dangelo2020} show that an evaluation at the level of the individual is reliable only when it rests on high-quality publication records, and what is required to assemble them. Performance remains strongly conditioned by name commonality, metadata completeness and corpus composition. The methods literature evaluates technique, usually against curated collections of modest size; the accuracy of the assignments that a deployed source actually publishes, at the scale at which downstream systems consume them, has received far less attention.

\textbf{ORCID and the limits of persistent identifiers.} ORCID places the record attached to each researcher identifier under the researcher's own control \citep{haak2012}. Publications accumulate through a mixture of self-assertion and consented automation, as publishers collect identifiers during submission and metadata services propagate new publications under standing permissions, so that the person, rather than an algorithm, states what they wrote. Engagement is uneven and measurably conditioned by local research environments and national infrastructure \citep{porter2022orcid}, so an ORCID record cannot be assumed complete, although exact agreement between an ORCID record and a source's disambiguation supplies an independent signal of quality. Persistent identifiers make curation lighter for every use, but they do so by attaching outputs unambiguously to a named person. Some researchers decline an identifier, or keep their record private, out of concern for privacy and data collection \citep{porter2022orcid,porterSR2025}; in the European Union the processing of personal data in a CRIS falls under the General Data Protection Regulation \citep{gdpr2016}; and the integration of many data sources in a CRIS has been noted to make anonymity increasingly hard to ensure when CRIS data are used for research \citep{schopfel2020ethics}.

\textbf{Identity in integrity signals.}
Research-integrity tools measure behaviour rather than documents, and behaviour belongs to people. The detection of implausibly prolific authorship begins from a count of one person's papers \citep{ioannidis2018}, that of authorship-for-sale operations from the shape of one person's collaboration network \citep{porter2024networks}, and that of citation cartels from citation between individuals \citep{fister2016}, at a time when the fabrication of papers has itself become industrialised \citep{else2021,cope2022}. Author identity is not itself an integrity signal, but many integrity signals are computed from them, with significant consequences if action is taken based on incorrect assertions.

%% file: sections/03_methods.tex
\section{Data and methods}\label{sec:methods}

\begin{figure*}[t]
\centering
\includegraphics[width=\textwidth]{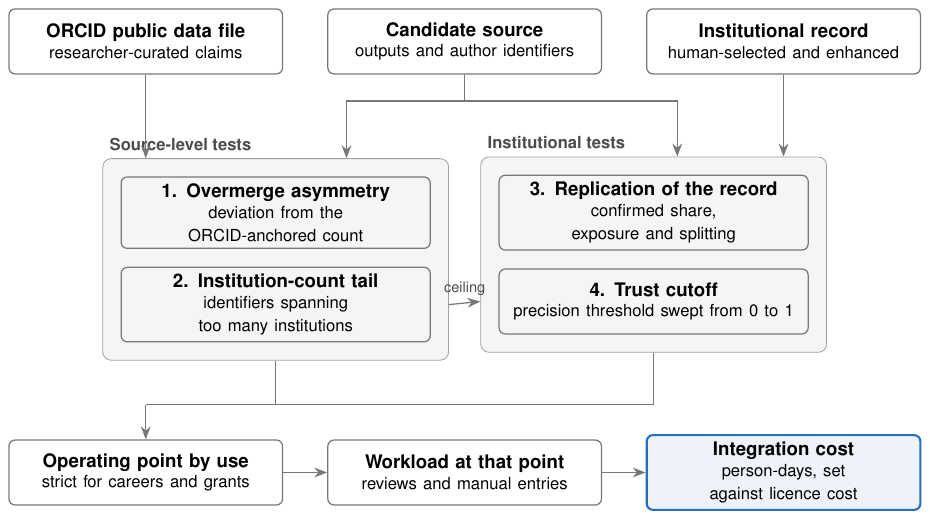}
\caption{Outline of the method. The source-level tests require no institutional data; the institutional tests read the candidate source against the institution's curated record. The institution-count test also supplies the conflation ceiling applied in the trust cutoff. The operating point is chosen according to the use to which the data are put, and the workload it implies is expressed in person-days.}
\label{fig:method}
\end{figure*}

The method requires three inputs: the ORCID public data file, the research outputs and author identifiers of a candidate source, and an institution's own confirmed publication--researcher records. Any institution holding these can apply the four tests to any candidate source without privileged access to it (Figure~\ref{fig:method}). The tests rest on ground truths that make different assumptions, and they read the direction of a deviation as well as its size, since some failures can move a count only one way. Results are reported as curves across a decision threshold, leaving the operating point to the institution, and are expressed throughout in person-days of research-office time.

\subsection{Sources and benchmark}\label{sec:data}

Five datasets are joined in Google BigQuery. OpenAlex is used through a complete relational mirror of its research outputs and authors \citep{priem2022}, and Dimensions through its publications and researchers tables \citep{herzog2020,dimensions_bq_docs}. ORCID records come from the public data file of researcher-curated records, in the copy hosted on BigQuery in a partnership between ORCID and Digital Science \citep{haak2012,orcid2026bq}. Crossref metadata holds the document universe constant \citep{hendricks2020}: OpenAlex also indexes DataCite DOIs, for example, so every count is restricted to Crossref-registered journal articles in order that comparisons are made over the same documents rather than reflecting differences in registry coverage. Both OpenAlex and Dimensions revise their author disambiguation continuously, so every figure reported here describes the sources as observed on \srcAnalysisDate{}: the OpenAlex release of \srcOpenAlexRelease{}, the Dimensions release of \srcDimensionsRelease{}, the Crossref release of \srcCrossrefRelease{}, the 2026 ORCID public data file (loaded \srcOrcidLoaded{}) and the Discovery records as last seen on the portals on \srcDiscoveryHarvest{}.

\paragraph{Choice of sources.} OpenAlex and Dimensions are compared because they are constructed in the same way. Both are built on open persistent-identifier infrastructure and openly available metadata, with Crossref and ORCID among their foundations, both assign author identities algorithmically, and openness was a stated aim in the design of each \citep{priem2022,hook2018}. Scopus and Web of Science follow a more editorially curated model, both in what they index and in how author profiles are maintained, so that a comparison across the two models would confound the way a source is built with the quality of its disambiguation. The method itself applies to any source.

\paragraph{The institutional benchmark.} Institutional ground truth is taken from the public research-profile portals of institutions that run Symplectic Elements, referred to here as \emph{Discovery} instances. Discovery is not a single system: each instance belongs to one institution and reflects that institution's own sources, settings and curation practice, and the study draws on \statNAll{} of them. Each instance is compared only with the researchers of its own institution, and \emph{Discovery} without qualification refers to the collection. Each exposes publication records, researcher records and the confirmed links between them, statements of the form ``this publication belongs to this member of staff''. How those links arise is central to their use as a benchmark. Each Elements instance harvests candidate publications from the data sources it has enabled: sources that need no institutional content licence, such as arXiv, DBLP, Europe PMC, PubMed, RePEc and SSRN (some through Symplectic's hosted data sources package), connected institutional repositories, and Crossref, which is used to look up records by their identifiers and also comes through that package, together with licensed sources that an institution may add, among them Web of Science, Scopus and Dimensions; OpenAlex is not among the documented sources \citep{symplectic_datasources}. Candidates are found by searching each source with the researcher's name variants and address terms. For arXiv, Dimensions, Scopus, SSRN and Web of Science the search can also use that source's own author identifier, combined with the name search, and a claimed ORCID iD retrieves records from the sources that carry it in their metadata \citep{symplectic_searchsettings}. A candidate enters a researcher's record only when the researcher, or a colleague acting on their behalf, claims it, except where the researcher has marked an identifier, whether an ORCID iD or a source's author identifier, as always identifying them, in which case matching items are claimed automatically \citep{symplectic_autoclaim}. The institution sets a default order of preference among the sources' versions of each output, which can be overridden for an individual output \citep{symplectic_preferred}; a manual record can be added to correct an output's metadata, and records can be added that no source holds. The result is a human-confirmed and human-enhanced record of authorship, maintained for operational rather than evaluative purposes, which draws on a source's disambiguation only where a researcher has chosen to search or claim by that source's identifier (Section~\ref{sec:limitations}), and we treat it as a proxy for a golden dataset. Because the comparison is joined on DOI, an institution's choice of preferred metadata affects how a record is displayed but not whether a person--publication link exists. The record is nonetheless incomplete: a source attributing a publication that the institution's Discovery instance lacks may be identifying a genuine gap or may have made a false match, and the analysis is designed around that ambiguity (Section~\ref{sec:insttests}).

\paragraph{The institutional sample.} The records were obtained from publicly available Discovery profile information, a convenience sample rather than the full Elements customer base. Of the instances accessible to the study, 55 carried enough journal-article history to be scored on a year-on-year currency ratio. An instance whose latest year shows a marked fall in recorded publications cannot serve as ground truth for recent output, whatever the cause, so an instance is included only if its 2025 journal-article count exceeds 90\% of its 2024 count; this retains \statNAll{} of the 55, one of the four excluded being an internal test account. The primary analysis is further restricted to the \statNPrimary{} institutions with more than 1,000 Discovery user profiles, since the smallest instances, some contributing fewer than a hundred matched researchers, produce visibly noisier ratios. The full set of \statNAll{} is retained as a robustness comparison. Of the \statNAll{} instances, \countryAllUS{} are in the United States, \countryAllGB{} in the United Kingdom, \countryAllAU{} in Australia, \countryAllNZ{} each in New Zealand, Canada and Germany, \countryAllIE{} in Ireland, one each in Saudi Arabia, South Africa and the United Arab Emirates, and one is an international research network; \countryAngloAll{} of the \statNAll{}, and \countryAngloPrimary{} of the \statNPrimary{} in the primary analysis, sit in majority-anglophone research systems.

\subsection{Source-level tests}\label{sec:sourcetests}

\paragraph{Overmerge asymmetry.} An author's true publication set is never observed directly, but when a self-curated ORCID record and a source's own disambiguation arrive at exactly the same publication count for a person, that agreement is reasonable evidence that the profile is complete and correct. We use such an agreeing pair as an anchor and measure the third source against it: Comparison~1 holds ORCID and Dimensions in agreement and measures OpenAlex, and Comparison~2 holds ORCID and OpenAlex in agreement and measures Dimensions. The anchor is strict, in that the anchoring source's profile must contain exactly the DOIs that the ORCID record asserts. Counts are taken over a common universe of DOIs indexed by Dimensions with publication years up to 2025, the last complete year in the annual ORCID public data file, so that researchers have had time to add their works to their records; an ORCID record enters only if it asserts more than five DOIs within that universe; and the profile in each source must carry the ORCID iD, since the comparison joins through it.

Merging two people into one profile can only add records to it, so a source that overmerges accumulates its deviations on the ``more than agreed'' side, whereas ordinary error misses and overshoots in comparable measure. We therefore report the share of authors under, equal to and over the agreed count, together with the mean signed deviation. The two comparisons run over different populations by construction, so the test compares the shapes of two distributions rather than pairing authors. An ORCID record could also map to several identifiers in the measured source, the opposite failure to overmerging, which does not inflate the ``over'' share: \fragOAMultiOrcids{} of \fragOAOrcids{} anchored ORCID records (\fragOAMultiOrcidsPct\%, accounting for \fragOARowsMultiPct\% of rows) map to more than one OpenAlex identifier, and \fragDimMultiOrcids{} of \fragDimOrcids{} to more than one Dimensions identifier. The anchored totals reported in Section~\ref{sec:overmerge} count ORCID--identifier pairs rather than ORCID records, and so exceed these counts by the additional pairs of the records that map to more than one identifier.

\paragraph{The institution-count tail.} Over a career a researcher accumulates a handful of institutional affiliations, not dozens, so the distribution of distinct institutions attributed to one author identifier is a direct reading of conflation. Both sources are first cleaned of cross-product records, on which every author is tagged with every institution on the paper, as this is a common enough error in publisher records to skew the result; these are detected where a paper carries more than three organisations and the mean number of institutions per author position equals the total on the paper. Two thresholds are defined for later use: an identifier spanning more than 40 institutions is treated as conflated and excluded from the trust cutoff, and a stricter ceiling of 20 defines a sensitivity check.

\subsection{Institutional tests}\label{sec:insttests}

Matching between Discovery users and source author identifiers favours precision over recall. A candidate match requires the DOI together with a surname test, extended to the full name only where the surname alone cannot distinguish between a paper's authors, and any case that still resolves to more than one candidate is discarded rather than guessed. A second strand matches directly on ORCID where the Discovery user has one recorded. Where two of a researcher's identifiers appear on the same paper, so that the researcher seems to collaborate with themselves, the researcher is removed from that source's matched population altogether, not just the two identifiers. Two authors of one paper cannot be the same person, so at least one of the two identifiers belongs to someone else, usually a co-author with the same surname. The rule is the same for both sources. In the primary sample it removes \statOASelfCollabRemoved{} of the \statOASelfCollabN{} researchers with candidate OpenAlex identifiers (\statOASelfCollabPct\%) and \statDimSelfCollabRemoved{} of the \statDimSelfCollabN{} with candidate Dimensions identifiers (\statDimSelfCollabPct\%). These are researchers whose names are hard to tell apart from a co-author's. To see how reliable their records are, we repeated the matching with the rule reversed, keeping only the researchers it would remove, and computed the same measures for them as for everyone else. Their records prove less reliable. They are prolific, with \statOASelfCollabWorksPer{} attributed works each in OpenAlex against \statOAKeptWorksPer{} for the researchers retained, and \statDimSelfCollabWorksPer{} against \statDimKeptWorksPer{} in Dimensions, so that they hold \statOASelfCollabWorksSharePct\% and \statDimSelfCollabWorksSharePct\% of each source's attributed works. A smaller share of those works is confirmed by their institutions: \statOASelfCollabConfPct\% against \statOAKeptConfPct\% in OpenAlex, and \statDimSelfCollabConfPct\% against \statDimKeptConfPct\% in Dimensions. Leaving them out therefore raises both sources' confirmed shares. Including them would lower the shares to \statOAWithRemovedConfPct\% and \statDimWithRemovedConfPct\%, which widens the gap between the sources slightly.

\paragraph{Replication of the record.} The \emph{new overlap} is the set of outputs that the source attributes to a researcher and that exist in the publication table of the institution's Discovery instance but are not linked to that person: low-risk additions, since the institution already holds the paper. The \emph{unconfirmed exposure} is the set of outputs attributed to the researcher whose publication appears nowhere in the institution's Discovery instance. Both are counted as attribution assertions, once per person, which is the appropriate unit for curation since each person--publication claim is reviewed separately. The exposure therefore mixes two things: publications the institution has not yet recorded, and outputs wrongly attributed to its researchers. The same match also measures splitting, the share of matched researchers whose outputs a source scatters across more than one identifier.

\paragraph{The trust cutoff.} Each candidate identifier matched to a Discovery user carries a precision ratio, the share of the source's attributed outputs that the institution's Discovery instance confirms. Sweeping a threshold on that ratio from 0 to 1 traces the trade-off: at a permissive threshold a CRIS accepts more researchers and recovers more of its record but admits more unconfirmed exposure, and as the threshold tightens precision rises and recall falls. Identifiers beyond the conflation ceiling are excluded beforehand, and only the best-matching identifier is kept for each researcher. A researcher's \emph{best match} is the identifier under which the institution confirms the most outputs, with ties broken by the higher precision ratio; it is chosen once, before the threshold is applied. The resulting curve is the integration-cost curve for that source at that institution, and the operating point on it is a property of the use: coverage-oriented reporting, where a missed publication is the costlier error, sits at a permissive threshold, and promotion cases or integrity-adjacent work at a strict one. Where a single point is needed for comparison we adopt 0.5, accepting an identifier only if the institution confirms at least half of its attributed outputs. The effort model below gives a consistent result. At its central unit times an identifier is worth delivering only when at least about half of its outputs are confirmed, so administrator effort is lowest at 0.5 for both sources. That point moves with the unit times, from \effOptThresholdInvestLow{} when investigation is quick to \effOptThresholdInvestHigh{} when it takes a quarter of an hour (Section~\ref{sec:cost}).

\subsection{Integration effort}\label{sec:effortmethod}

Integration effort is expressed in person-days of \effMinutesPerDay{} minutes, a working day of seven and a quarter hours; no monetary value is attached. The model counts only the time spent deciding attributions: verifying, checking, investigating and entering outputs. It leaves out the rest of the work of keeping a publication record, such as correcting metadata, checking open-access compliance, following up with researchers and monitoring the feed itself, for example checking that a source's identifiers have not drifted to point at different people as its algorithms are revised \citep{porter2024barcelona}. Its figures are therefore a floor on the effort a source imposes, not an estimate of the staff a research office needs, and sources are best compared through their ratios. The model also assumes that an administrator keeps the same pace through any number of decisions, with no loss of accuracy. That is unreasonable for a large backlog: the quality of repeated judgements falls over a long session, as when clinicians prescribe antibiotics more often as a clinic session goes on \citep{linder2014time} and radiologists detect abnormalities less accurately at the end of a long working day \citep{krupinski2010long}. A research office working through a backlog must either slow down or accept more errors, so the model probably undercounts, substantially, the time needed to process every decision well. Two workflows are modelled. In the first an administrator in the research office reviews the feed in batches; in the second researchers review the suggestions about their own work in response to email notifications, sent daily or weekly. Throughout, \emph{research office} stands for whichever central service curates the institution's record. In many institutions this work is carried out or supported by the research library, and the administrator's effort modelled here falls equally on library staff. In both the feed is treated as the institution's only source of candidate records, so that any confirmed output it fails to deliver must be entered by hand. In practice a CRIS may harvest several sources at once, which lowers this term; the single-feed assumption isolates the contribution of the source being assessed.

Four activities are distinguished. An output that the feed delivers and the institution confirms has to be \emph{verified}. An output that the feed delivers but the institution has not confirmed has to be \emph{checked}, and \emph{rejected} if it proves not to belong to the researcher. An output that the institution confirms but the feed does not deliver has to be \emph{entered} by hand.

Table~\ref{tab:effortmodel} sets out what each activity costs an administrator and a researcher. The costs differ in two ways.

For an administrator, an unconfirmed output takes longer than a confirmed one, and how much longer depends on the identifier that carries it. If the institution confirms most of that identifier's outputs, an unconfirmed one is probably a genuine publication that has not yet been recorded, requiring only a quick check ($t_c$). If the institution rarely confirms the identifier's outputs, the unconfirmed one probably belongs to someone else, and showing that takes a full investigation ($t_x$). The model blends the two according to the identifier's precision ratio $r$. An unconfirmed output costs $t_c + (1-r)(t_x-t_c)$: a quick check when $r=1$, a full investigation when $r=0$, and in between in proportion. Identifiers beyond the conflation ceiling, and identifiers other than the researcher's best match (Section~\ref{sec:insttests}), are given $r=0$. Neither kind is delivered once a threshold is applied, so this matters only for unfiltered ingestion, where every identifier matched to a researcher reaches the feed. 

Researchers recognise their own work, so every decision, whether to claim or to reject, takes them the same short time $t_d$, whatever the identifier. Instead they pay an overhead $s$ for each session in which they respond to a notification: logging in, authenticating and returning to the task they left. A researcher logs in because they have published, whichever source feeds the CRIS: each genuine new output brings them in once in its notification period, a day or a week, to confirm a suggestion or to enter the output by hand. Sessions are therefore counted for each researcher and period with at least one genuine new output, expected from the outputs either source finds and so the same for both sources, plus the periods in which a source sends only suggestions that prove unconfirmed, when the researcher logs in only to reject them. An output that has to be entered takes the same time $t_e$ whoever enters it.

\begin{table*}[t]
\centering
\small
\caption{What each activity costs an administrator and a researcher. Total effort is the sum, over all outputs, of the cost of the activity each output needs, plus the session overheads.}
\label{tab:effortmodel}
\begin{tabular}{llll}
\toprule
Activity & Number & Administrator & Researcher\\
\midrule
Verify a delivered, confirmed output & $n_v$ & $t_v$ & $t_d$\\
Check, and if need be reject, a delivered, unconfirmed output $u$ & $n_u$ & $t_c + (1-r_u)(t_x-t_c)$ & $t_d$\\
Enter a confirmed output the feed misses & $n_e$ & $t_e$ & $t_e$\\
Respond to a notification & $k$ & --- & $s$\\
\bottomrule
\end{tabular}
\end{table*}

Adding up the rows of Table~\ref{tab:effortmodel} gives the effort of each workflow:
\begin{align}
T_{\mathrm{admin}} &= n_{v}\,t_{v} + \sum_{u=1}^{n_u}\bigl[t_{c} + (1-r_u)\,(t_{x}-t_{c})\bigr] + n_{e}\,t_{e}, \label{eq:admin}\\
T_{\mathrm{res}} &= n_{v}\,t_{d} + \sum_{u=1}^{n_u} t_{d} + n_{e}\,t_{e} + k\,s. \label{eq:res}
\end{align}
Both sums run over the same $n_u$ unconfirmed outputs. For an administrator each term is different, because each output's identifier has its own precision ratio $r_u$. For a researcher every term is $t_d$, so the sum is simply $n_u t_d$ and $T_{\mathrm{res}} = (n_v + n_u)\,t_d + n_e\,t_e + k\,s$. Table~\ref{tab:effortparams} gives the unit times.

\begin{table*}[t]
\centering
\small
\caption{Unit times of the effort model. Central values are used unless stated; the range gives the low and high settings. Two parameters, the time to investigate an attribution ($t_x$) and the overhead of a researcher's session ($s$), are not measured in any published source: both are varied across their ranges.}
\label{tab:effortparams}
\begin{tabular}{lrrl}
\toprule
Activity & Central & Range & Basis\\
\midrule
Verify a confirmed output, $t_v$ & \effVerifySec{}\,s & \effVerifySecLow--\effVerifySecHigh\,s & screening medians \citep{oblenis2017vote}\\
Check an unconfirmed output, $t_c$ & \effCheckSec{}\,s & \effCheckSec--\effCheckSecHigh\,s & screening mean \citep{howard2020swift}; copy cataloguing \citep{siguenzaguzman2014cataloguing}\\
Investigate an attribution, $t_x$ & \effRejectMin{}\,min & \effRejectMinLow--\effRejectMinHigh\,min & not measured; upper value from full-text assessment \citep{trad2025streamlining}\\
Enter a missing output, $t_e$ & \effEntryMin{}\,min & \effEntryMin--\effEntryMinHigh\,min & manual entry \citep{fccn2024cienciavitae,symplectic2016qmul}; deposit \citep{ua_ir_deposit}\\
Researcher decision, $t_d$ & \effResDecideSec{}\,s & & as $t_v$\\
Session overhead, $s$ & \effOverheadMin{}\,min & 0--\effOverheadMax\,min & not measured; varied; authentication \citep{reynolds2020systemic}\\
\bottomrule
\end{tabular}
\end{table*}
Effort is reported for two periods. The first is the \emph{backlog at adoption}: the one-off effort of bringing an institution's existing record into line with a source when the source is first connected. It covers every output that had reached the source before \effWindowStart{}, and uses the confirmations in the Discovery instances as they were observed.

The second is the \emph{recurring effort}: the work that a connected source generates each year as new outputs arrive. It covers the twelve months from \effWindowStart{} to \effWindowEnd{}, the most recent full year before the observation date. A feed delivers an output when the output is added to the source, not when it was published, so the recurring effort is measured by the date each output was added. Dimensions records that date for every publication \citep{dimensions_bq_docs}, and its index is updated daily \citep{hook2021realtime}. OpenAlex's record creation dates cannot be used, because they were reset in bulk in October 2025. Each output is therefore dated by the day the same DOI entered Dimensions, for both sources. Dimensions holds \effOADimClockPct\% of the DOIs that OpenAlex attributes; the rest are dated by their Crossref registration.

Two quantities in the recurring effort cannot be read directly from the data. The first is whether a new output will be confirmed. For this investigation, the Discovery instances were last observed on \srcDiscoveryHarvest{}, so outputs added during the year had little or no time to be confirmed before it, and none after. Each new output is instead given a probability of confirmation equal to the precision ratio of its identifier before the year began. The second is the number of outputs that will need to be entered by hand. That depends on how much the researchers publish, which is the same whichever source the institution uses, so a single publication rate is used for both. It is estimated in two steps: the expected number of genuine new outputs found by either source in the year, divided by the share of the institutional record that the two sources together recover. The result is some \effTrueRate{} outputs a year across the primary sample.

Two summaries are reported. For the backlog at adoption, which a research office takes on when it connects a source, the threshold at which administrator effort is lowest is used. For the recurring effort, the researcher workflow: only researchers can tell that an identifier has stopped carrying their work because another has been created, and only they control the ORCID route by which new outputs are claimed automatically. It is reported as the difference between the sources in person-days and as their ratio, with the session overhead and the entry time varied together. The two act differently. Every session is shared by both sources, so a longer overhead adds the same time to each and narrows their ratio, whereas entry falls mostly on the source that misses more output, so a longer entry time widens the gap. A low overhead with slow entry therefore gives a different comparison from a high overhead with fast entry, even where the totals for one source are similar.

%% file: sections/04_results.tex
\section{Results}\label{sec:results}

\subsection{Replicating the institutional record}\label{sec:replication}

Pooled across the \statNPrimary{} primary-scope institutions, the two sources reconstruct the institutions' own records to markedly different extents (Table~\ref{tab:replication}). Of the publications that OpenAlex attributes to matched researchers, \statOAConfirmedPctPrimary\% are confirmed by an existing publication--researcher link in the institution's Discovery instance; for Dimensions the confirmed share is \statDimConfirmedPctPrimary\%. Read from the side of risk, \statOANotInSystemPctPrimary\% of OpenAlex's attributed research outputs are assertions whose publication is entirely absent from the institution's own record, against \statDimNotInSystemPctPrimary\% for Dimensions, an exposure \statExposureRatio{} times as large as a share of each source's own claims. Both sources also surface low-risk additions, publications that the institution holds but had not yet linked to the person. The larger exposure indicates how much of a source's output arrives unverifiable, though not that every such attribution is wrong. These pooled shares barely depend on which institutions are included: recomputed over the full set of \statNAll{}, the confirmed shares are \statOAConfirmedPctAll\% and \statDimConfirmedPctAll\%, so the small institutions add noise to per-institution ratios but not bias to the pooled totals.

\begin{table*}[t]
\centering
\small
\caption{Pooled replication of institutional records, by source and scope. The institutional baselines (confirmed journal-article statements in the Discovery instances) are \repBaselinePrimary{} publications for the primary scope and \repBaselineFull{} for the full scope. \emph{New overlap}: research outputs that the source attributes and the institution holds but had not linked to the person. \emph{Unconfirmed exposure}: attribution assertions whose publication is absent from the institution's record.}
\label{tab:replication}
\begin{tabular}{lrrrr}
\toprule
 & \multicolumn{2}{c}{Primary scope ($n=\statNPrimary$)} & \multicolumn{2}{c}{Full scope ($n=\statNAll$)}\\
\cmidrule(lr){2-3}\cmidrule(lr){4-5}
 & OpenAlex & Dimensions & OpenAlex & Dimensions\\
\midrule
\input{generated/replication_rows.tex} 
\bottomrule
\end{tabular}
\end{table*}

\subsection{The trust cutoff}\label{sec:trust}

At every precision threshold from 0 to 1, pooled across institutions, OpenAlex retains fewer trusted researcher profiles than Dimensions, and the gap is widest at the moderate thresholds that an adopting institution is most likely to choose. At a threshold of 0.5 the retention ratio, OpenAlex users retained per Dimensions user retained, has a primary-scope median of \statMedianRatioPrimary; at the typical institution, in other words, Dimensions yields about \statDimMoreProfilesPct\% more researcher profiles that meet the threshold.

\begin{figure*}[t]
\centering
\includegraphics[width=\textwidth]{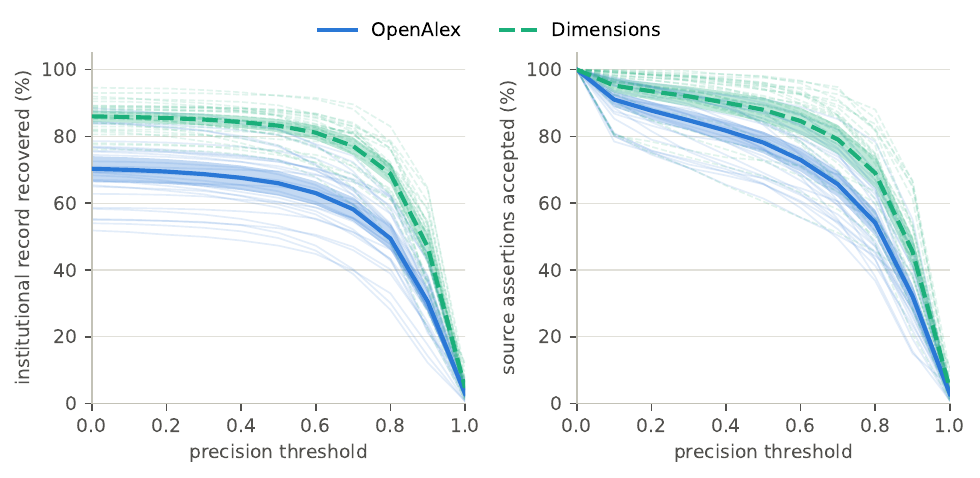}
\caption{The trust-cutoff sweep across the 30 primary-scope institutions. One faint trace per institution; bold lines are unweighted means with shaded 95\% confidence intervals, disjoint at every threshold from \statRecallDisjointFrom{} (left) and \statRetentionDisjointFrom{} (right). Left: the institution's own confirmed record recovered, as a percentage of that record. Right: the source's attributed research outputs accepted, as a percentage of the source's value at threshold zero. Both panels use the conflation ceiling of 40.}
\label{fig:trust}
\end{figure*}

Figure~\ref{fig:trust} traces the full curves across the \statNPrimary{} institutions of the primary sample, normalised so that institutions of very different sizes share one axis. On the recall side, Dimensions recovers on average \statDimRecallPermissive\% of an institution's own record at a permissive threshold, against \statOARecallPermissive\% for OpenAlex, and holds that advantage as the threshold tightens. On the retention side the sharpest feature is the first step: \statOAFirstStepPct\% of OpenAlex's assertions are removed as soon as a tenth of an identifier's attributed outputs must be corroborated, since they belong to identifiers whose outputs the institution essentially never confirms, whereas Dimensions loses \statDimFirstStepPct\%. The spread between institutions is far smaller than the gap between the sources: the 95\% confidence bands around the two means are disjoint on the recall panel at every threshold from \statRecallDisjointFrom{} and on the retention panel at every threshold from \statRetentionDisjointFrom{}, so the separation holds at every threshold an institution would realistically choose.

\begin{figure}[t]
\centering
\includegraphics[width=0.9\columnwidth]{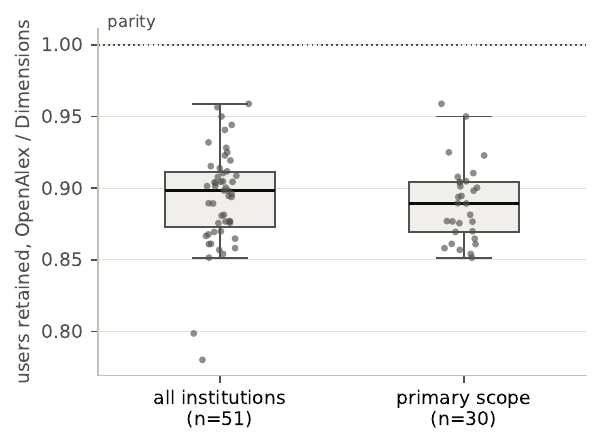}
\caption{Per-institution retention ratio (OpenAlex users retained per Dimensions user retained) at a precision threshold of 0.5. Each point is one institution.}
\label{fig:retention}
\end{figure}

Although the \statNPrimary{} institutions differ considerably in size, disciplinary profile and curation practice, the retention ratio falls below one at every one of them (Figure~\ref{fig:retention}), within a narrow band (\statRatioMinPrimary--\statRatioMaxPrimary, standard deviation \statStdPrimary). On the full set of \statNAll{} institutions, including the smallest instances excluded by the size floor of the primary scope, the ratio falls below one in \statRatioBelowOneAll{} of them.

\subsection{What filtering removes}\label{sec:decomposition}

\begin{figure*}[t]
\centering
\includegraphics[width=\textwidth]{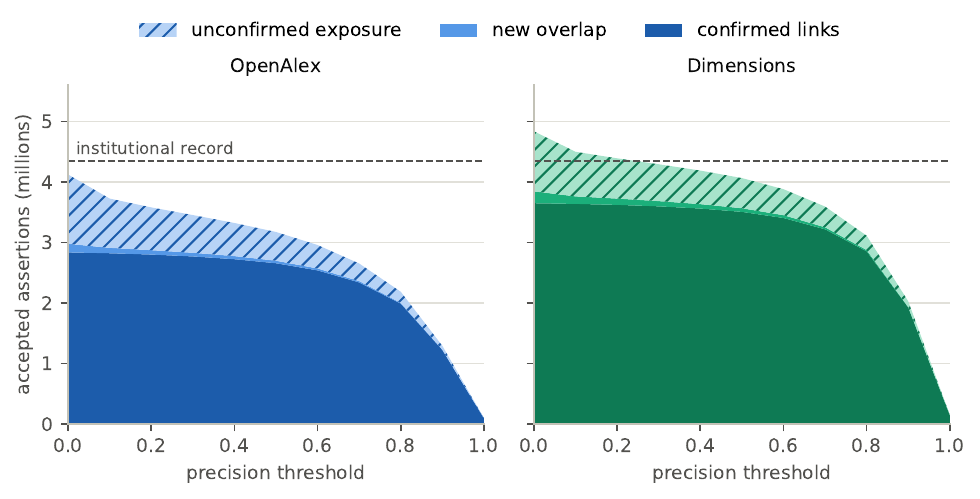}
\caption{Decomposition of accepted assertions under the trust cutoff, pooled across the \statNPrimary{} primary-scope institutions (conflation ceiling 40). Bands, from bottom to top within each panel: confirmed links, new overlap and unconfirmed exposure (hatched). The dashed line marks the pooled institutional record. These populations sit below those of Table~\ref{tab:replication}, which applies neither the conflation ceiling nor best-identifier selection.}
\label{fig:decomposition}
\end{figure*}

The decomposition of Table~\ref{tab:replication} can be recomputed at each threshold. Figure~\ref{fig:decomposition} splits the assertions accepted at each threshold into confirmed links, new overlap and unconfirmed exposure, pooled across the primary sample. For both sources most of what is filtered out as the threshold tightens is exposure: between thresholds of 0 and 0.5, \statOAExpRemovedPct\% of OpenAlex's unconfirmed exposure is removed at a cost of \statOAConfLostPct\% of its confirmed content, while for Dimensions the corresponding figures are \statDimExpRemovedPct\% and \statDimConfLostPct\%. The gap between the sources nonetheless remains at every operating point. The accepted OpenAlex feed carries both a smaller confirmed base and, until the far right of the range, a larger share at risk than the accepted Dimensions feed, whose total tracks the institutional record through the middle of the range.

\subsection{Why the sources differ: Lumping and splitting}\label{sec:overmerge}

\begin{figure}[t]
\centering
\includegraphics[width=0.8\columnwidth]{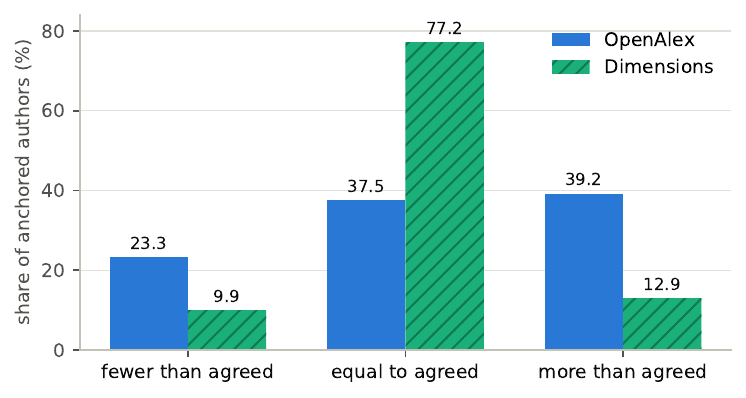}
\caption{Deviation from the agreed publication count, by measured source. Each source is measured against the anchor formed by exact agreement between ORCID and the \emph{other} source (Section~\ref{sec:sourcetests}); the populations therefore differ, and the signal is the shape of each distribution rather than its level. Overmerging can only add publications to a profile, so it accumulates in the rightmost pair.}
\label{fig:overmerge}
\end{figure}

Anchored on exact agreement between ORCID and Dimensions, OpenAlex deviates from the agreed publication count asymmetrically (Figure~\ref{fig:overmerge}): \statOAOverPct\% of the \statOANAuthors{} anchored authors carry more publications than the agreed set, with a mean excess of \statOAMeanDev{} publications, while \statOAUnderPct\% carry fewer. Measured against the mirror anchor of agreement between ORCID and OpenAlex, Dimensions shows the balanced profile that random error predicts: \statDimOverPct\% over, \statDimUnderPct\% under and \statDimEqualPct\% of its \statDimNAuthors{} anchored authors in exact agreement, with a mean excess of \statDimMeanDev. The two populations differ by construction, so the raw counts are not comparable across them; what matters is the shape of each distribution: Dimensions' deviations are spread evenly above and below the agreed count, whereas OpenAlex's are concentrated above it, which is what merging profiles would produce.

\begin{figure}[t]
\centering
\includegraphics[width=0.95\columnwidth]{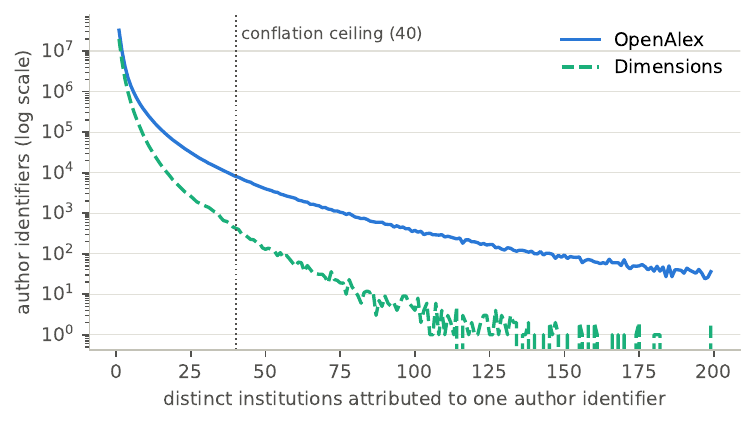}
\caption{Distinct institutions attributed to a single author identifier, after removal of cross-product affiliation records; note the logarithmic scale. Identifiers beyond the dotted conflation ceiling are excluded from the trust-cutoff analysis. The axis is truncated at 200 institutions; the OpenAlex distribution continues beyond it.}
\label{fig:insttail}
\end{figure}

The conflation signature appears again, independently, in the affiliation data (Figure~\ref{fig:insttail}). After the removal of cross-product affiliation records, \statOATailN{} OpenAlex author identifiers span more than 40 distinct institutions, a figure with no plausible explanation in terms of a single career, against \statDimTailN{} Dimensions researcher identifiers, a ratio of roughly \statTailRatio{} to one. Each identifier in that tail is one that a CRIS would have to quarantine or resolve by hand, so the tail expresses part of the integration cost as a count.

The same institutional match exposes the opposite failure. Of the researchers matched to a Discovery user across the \statNPrimary{} primary-scope institutions, OpenAlex assigns \statOAMultiIdPct\% to more than one author identifier, against \statDimMultiIdPct\% for Dimensions, and OpenAlex is the more fragmented in every one of the \statNPrimary{} institutions. The sources thus differ on both failure modes, each of which inflates the curation an institution inherits, since a split profile must be reassembled just as a lumped one must be pulled apart. Because OpenAlex assigns an identifier to every author on every work, \statOASinglePubPct\% of its author identifiers across the Crossref journal-article record attach to only one publication, against \statDimSinglePubPct\% for Dimensions, and its higher split rate might be expected to be an artefact of such identifiers. Removing them, and counting a researcher as fragmented only when at least two of their identifiers each carry more than one publication, lowers the split rates only marginally, to \statOAMultiIdPctFiltered\% for OpenAlex and \statDimMultiIdPctFiltered\% for Dimensions, and OpenAlex remains the more fragmented in every one of the \statNPrimary{} institutions. The splitting extends to new output. The test takes the researchers matched in both sources and, in one source, those of their identifiers with a reliable history: at least five works, at least 80\% of them confirmed by the institution. Of the \effNewWorksDimN{} works that Dimensions added under such identifiers between \effWindowStart{} and \effWindowEnd{}, OpenAlex attributes \effOAKeepsDimNewPct\% to the same researcher. Run the other way, of the \effNewWorksOAN{} works that OpenAlex added under its own reliable identifiers, Dimensions attributes \effDimKeepsOANewPct\% to the same researcher. Few of the works that OpenAlex misses are absent from it (\effOAMissAbsentPct\%). Most, \effOAMissNewIdPct\%, sit instead under an identifier that OpenAlex's own author record shows was created in or after November 2025. Earlier creation dates cannot be used, because OpenAlex reset most of them to a single day in October 2025, but the dates it has assigned since then mark when each identifier was made. Of those new identifiers,  \effOAMissNewIdSmallPct\% carry three works or fewer. A feed that follows a researcher's known OpenAlex identifiers therefore misses a larger share of what they publish next, because as measured OpenAlex often starts a new identifier for them.

\subsection{Sensitivity to the conflation ceiling}\label{sec:sensitivity}

The trust-cutoff analysis excludes identifiers spanning more than 40 institutions, and 40 is a judgement rather than a derived quantity. Rerun with the curation-grade ceiling of 20, pooled OpenAlex retention across the primary scope falls by \sensCeilingDropPct\% at the middle thresholds (\sensUsersTwentyPooled{} against \sensUsersFortyPooled{} researchers retained at a threshold of 0.5), that being the share of the retained population that had depended on identifiers in the band from 20 to 40 institutions, which a curation-grade feed would arguably exclude as borderline conflated. The qualitative conclusions of Section~\ref{sec:trust} are unchanged at either ceiling. What the check measures is how much apparent coverage rests on identifiers spanning 20 to 40 institutions, profiles whose breadth already suggests conflation although they fall below the formal ceiling.

\subsection{Integration effort}\label{sec:cost}

\begin{figure*}[t]
\centering
\includegraphics[width=\textwidth]{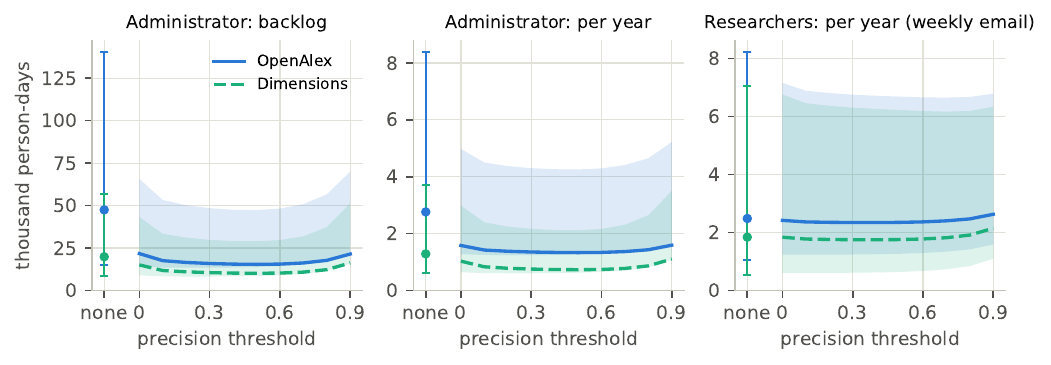}
\caption{Integration effort against the precision threshold, pooled across the \statNPrimary{} primary-scope institutions (conflation ceiling 40, best identifier); the point at the left of each panel is unfiltered ingestion. Lines are central estimates. Shading and error bars span the low and high parameters of Table~\ref{tab:effortparams} for the administrator, and session overheads from 0 to \effOverheadMax{} minutes about a central \effOverheadMin{} for researchers. Left: the backlog at adoption. Centre and right: effort in the year from \effWindowStart{}.}
\label{fig:effort}
\end{figure*}

Figure~\ref{fig:effort} and Table~\ref{tab:effort} carry the institutional tests through to effort under the model of Section~\ref{sec:effortmethod}. For an administrator taking on the backlog at adoption, unfiltered ingestion of OpenAlex requires some \effOABacklogNone{} person-days across the \statNPrimary{} institutions at the central parameters, against \effDimBacklogNone{} for Dimensions, \effBacklogRatioNone{} times as much. Most of the difference lies in investigation. Of OpenAlex's unconfirmed attributions, \effOAUnconfWeakIdPct\% sit under identifiers that are conflated or are not the researcher's best match, against \effDimUnconfWeakIdPct\% for Dimensions, and these are the attributions slowest to disprove. Filtering removes them first, and administrator effort falls to its minimum at a threshold of \effOAOptThreshold{} for OpenAlex and \effDimOptThreshold{} for Dimensions, at \effOABacklogHalf{} and \effDimBacklogHalf{} person-days, \effOAFilterSavingPct\% and \effDimFilterSavingPct\% below unfiltered ingestion. The curve is nearly flat around the minimum, staying within 5\% of it between thresholds of \effOAFlatLow{} and \effOAFlatHigh{} for OpenAlex and of \effDimFlatLow{} and \effDimFlatHigh{} for Dimensions; beyond that range the manual entry of what the filter discards outweighs the investigation it saves. At the minimum, OpenAlex still requires \effBacklogRatioHalf{} times the effort of Dimensions. It should be noted that where the minimum falls is set by the unit times rather than by the source. An identifier is worth delivering when the time saved on entering its confirmed outputs exceeds the time spent checking and investigating its unconfirmed ones, and at the central times that holds once about half its outputs are confirmed. If an investigation takes \effRejectMinLow{} minute the minimum moves to \effOptThresholdInvestLow{}, and if it takes \effRejectMinHigh{} minutes to \effOptThresholdInvestHigh{}. At each setting the minimum is the same for both sources, in the backlog and in the recurring effort.

The recurring effort falls on researchers (Section~\ref{sec:effortmethod}). In the year from \effWindowStart{}, with the feed filtered at 0.5 and a session overhead of \effOverheadMin{} minutes, researchers across the primary sample would spend \effOAResFlowHalf{} person-days on OpenAlex against \effDimResFlowHalf{} on Dimensions: \effResFlowGapHalf{} person-days more, or \effResFlowRatioHalf{} times as much. The gap holds across the range of overheads, from \effResFlowGapHalfMax{} to \effResFlowGapHalfZero{} person-days, while the ratio narrows from \effResFlowRatioHalfZero{} at no overhead to \effResFlowRatioHalfMax{} at \effOverheadMax{} minutes, as the logins that both sources share come to dominate. For an administrator the same year's attribution decisions alone would take \effFlowRatioHalf{} times as long for OpenAlex. Most of the difference is entry, because OpenAlex splits new output as well as old: OpenAlex attributes to the same researcher \effOAKeepsDimNewPct\% of the new works that Dimensions adds under a researcher's reliable identifiers, whereas Dimensions attributes \effDimKeepsOANewPct\% of the new works that OpenAlex adds under its reliable identifiers (Section~\ref{sec:overmerge}).

The comparison therefore depends more on how long entry takes than on how long a login takes (Table~\ref{tab:resgrid}). With fast entry and a \effOverheadMax-minute overhead, OpenAlex costs researchers \effResGapFastEntryHighLogin{} person-days more, only \effResRatioFastEntryHighLogin{} times as much. With \effEntryMinHigh-minute entry and no overhead, it costs \effResGapSlowEntryNoLogin{} more, \effResRatioSlowEntryNoLogin{} times as much. Even with both at their highest, the gap is \effResGapSlowEntryHighLogin{} person-days. The ratio alone understates the difference whenever logins are slow, and the gap understates it whenever entry is slow, so both are reported.

\begin{table}[t]
\centering
\small
\caption{Researchers' recurring effort: the extra person-days a year that OpenAlex costs researchers across the primary sample, and the ratio of the two sources, by the time to enter a missing output ($t_e$) and the overhead of a session ($s$). Feed filtered at 0.5, weekly email, other times central; the central setting is $t_e=\effEntryMin$\,min, $s=\effOverheadMin$\,min.}
\label{tab:resgrid}
\input{generated/effort_res_grid.tex}
\end{table}

\begin{table*}[t]
\centering
\small
\caption{Integration effort in person-days, pooled across the \statNPrimary{} institutions of the primary sample, under unfiltered ingestion and at a precision threshold of 0.5 (conflation ceiling 40, best identifier). Administrator rows give the central estimate and, below it, the range across the low and high parameters of Table~\ref{tab:effortparams}. Researcher rows assume a weekly email and \effOverheadMin{} minutes of overhead per session. Recurring effort covers the year from \effWindowStart{}; a person-day is \effMinutesPerDay{} minutes.}
\label{tab:effort}
\begin{tabular}{lrrrr}
\toprule
 & \multicolumn{2}{c}{OpenAlex} & \multicolumn{2}{c}{Dimensions}\\
\cmidrule(lr){2-3}\cmidrule(lr){4-5}
 & Unfiltered & Threshold 0.5 & Unfiltered & Threshold 0.5\\
\midrule
\input{generated/effort_rows.tex} 
\bottomrule
\end{tabular}
\end{table*}

Researchers log in because they publish: \effOASessionsWeekly{} weekly sessions a year for each matched researcher with OpenAlex and \effDimSessionsWeekly{} with Dimensions, the small excess for OpenAlex being weeks in which it sends only suggestions that prove unconfirmed. Suggestions reach them slowly: \effOASuggestionsPerYear{} a year for each matched researcher from OpenAlex and \effDimSuggestionsPerYear{} from Dimensions, so that the median notification carries \effOAItemsPerSession{} suggestion and a weekly email saves only \effOADigestSavingPct\% of the sessions that a daily one would require. Filtering helps researchers little, since it saves rejections that take them seconds while adding entries that take minutes: unfiltered, their recurring effort is \effOAResFlowNone{} person-days for OpenAlex and \effDimResFlowNone{} for Dimensions.

The ranges are wide, because the time to investigate an attribution has not been measured, but at each of the low, central and high settings OpenAlex requires more effort than Dimensions at every threshold and for both workflows: unfiltered it calls for far more investigation, and at every threshold for more entry. At the high setting, appropriate where the people concerned are not known to the reviewer, the effort at a threshold of 0.5 is about \effOAHighFactor{} times the central estimate for OpenAlex and \effDimHighFactor{} times for Dimensions. Tightening the conflation ceiling to 20 raises the filtered backlog by \effOACeilTwentyRisePct\% for OpenAlex and \effDimCeilTwentyRisePct\% for Dimensions, as outputs under identifiers spanning 20 to 40 institutions move from review to entry, without changing the comparison. Filtering makes the effects of overmerging less visible but leaves them present, since the accepted feed still carries whatever conflation its identifiers contain.

%% file: generated/replication_rows.tex
Attributed works & 7,121,642 & 5,648,698 & 7,556,534 & 5,982,223\\
Confirmed by institution & 3,591,679 & 3,933,917 & 3,824,726 & 4,173,540\\
Confirmed share (\%) & 50.4 & 69.6 & 50.6 & 69.8\\
New overlap & 212,489 & 207,058 & 213,907 & 209,234\\
Unconfirmed exposure & 3,317,474 & 1,507,723 & 3,517,901 & 1,599,449\\
Unconfirmed share (\%) & 46.6 & 26.7 & 46.6 & 26.7\\

%% file: generated/effort_res_grid.tex
\begin{tabular}{lrrrr}
\toprule
 & \multicolumn{4}{c}{Session overhead, $s$ (min)}\\
\cmidrule(l){2-5}
Entry, $t_e$ (min) & 0 & 2 & 5 & 10\\
\midrule
\multicolumn{5}{l}{\emph{Extra person-days}}\\
3 & 625 & 592 & 544 & 462\\
5 & 1,082 & 1,050 & 1,001 & 920\\
10 & 2,227 & 2,195 & 2,146 & 2,065\\
\midrule
\multicolumn{5}{l}{\emph{Ratio, OpenAlex to Dimensions}}\\
3 & 1.98 & 1.34 & 1.16 & 1.07\\
5 & 2.12 & 1.50 & 1.27 & 1.14\\
10 & 2.24 & 1.75 & 1.47 & 1.28\\
\bottomrule
\end{tabular}

%% file: generated/effort_rows.tex
Administrator, backlog at adoption & 47,486 & 15,410 & 19,878 & 10,159\\
\quad\emph{range} & 15,269--140,323 & 13,815--47,381 & 8,637--56,857 & 8,545--29,008\\
Administrator, recurring per year & 2,763 & 1,333 & 1,287 & 732\\
\quad\emph{range} & 1,298--8,382 & 1,261--4,255 & 626--3,707 & 625--2,119\\
Researchers, backlog at adoption & 10,050 & 14,117 & 7,413 & 9,063\\
Researchers, recurring per year & 2,482 & 2,348 & 1,839 & 1,756\\

%% file: sections/05_discussion.tex
\section{Discussion}\label{sec:discussion}

Each of the four tests finds more attribution error in OpenAlex than in Dimensions. OpenAlex exceeds the ORCID-agreed publication count \statOverRatio{} times as often (\statOAOverPct\% of anchored authors against \statDimOverPct\%). It has about \statTailRatio{} times as many identifiers spanning more than 40 institutions. A smaller share of its attributions is confirmed by the institutions (\statOAConfirmedPctPrimary\% against \statDimConfirmedPctPrimary\%). And it retains fewer trusted profiles at a threshold of 0.5 at every one of the \statNPrimary{} primary institutions. The institutional result is also unusually uniform, but the more general finding concerns the method rather than the comparison. Disambiguation quality can be measured in advance, from data that an adopting institution already holds, by tests that any institution can reproduce against any source, and the measurement converts directly into a quantity that can be planned around: the volume of claims about the institution's own people that will need checking before a feed can be trusted.

\subsection{Identifier trust is not uniform}

Within a single source, author identifiers do not all merit the same trust, and the signals computed here allow trust to be encoded in tiers: ORCID-corroborated identifiers at the top, identifiers within a plausible institution-count band accepted with monitoring, and identifiers beyond the conflation ceiling quarantined pending human review. Because these tiers can be detected in advance, the methods applied in this paper, used proactively, can substantially reduce the effort of integrating a secondary source's assertions into the trusted assertions of an institutional record. (In practice, in a CRIS such as Symplectic Elements, filtering decisions are made for each researcher.)

The different levels of trust can also be used both within and outside CRIS workflows. A record assembled for a promotion committee would draw only on confirmed links and, at most, on ORCID-corroborated identifiers; a statutory return might accept the middle tier subject to sample-based audit; and an analysis of national trends could operate at a permissive threshold, provided that the source's errors had been shown, by a test such as the overmerge asymmetry, not to lean in one direction. Taking the filter seriously has two consequences. Sources are better compared by the number of profiles that survive the filter, which Section~\ref{sec:trust} measures, than by the number they offer; and what the filter discards is itself a cost, approximating the cleaning an institution would take on if it chose to repair the feed rather than quarantine its weaker identifiers.

The effort of verification is not uniform either, and depends on what the institution is trying to establish. Confirming that a known colleague wrote a paper sits at the low end of the ranges in Table~\ref{tab:effortparams}. Assessing researchers who are not known to the reviewer, as in research-integrity investigations or in the research-security screening of prospective collaborators, applicants and visitors, belongs at the high end: the reviewer cannot rely on recognising the person's work, a doubtful attribution has to be investigated rather than dismissed, and a conflation that goes unnoticed carries consequences of its own. For such tasks effort should be modelled at the upper end of the ranges, where it is about \effOAHighFactor{} times the central estimate for OpenAlex and \effDimHighFactor{} times for Dimensions, and it is there that the tiers of trust described above matter most.

\subsection{Who carries the integration effort}

Because the same record drives reporting, compliance, profiles and assessment simultaneously, an institution needs it to be both comprehensive and accurate and cannot simply trade one for the other on the trust-cutoff curve; what the curve makes explicit is the effort this imposes, and sources are properly compared on the total effort of reaching the required record. Filtering does not remove that effort but moves it. Up to the minimum of the effort curve a filter trades investigations, which are slow for an administrator, for entries; beyond it, entries dominate. Where the minimum falls depends on how long an investigation takes, which has not been measured, but it is the same for both sources at every setting, so the comparison between them does not depend on it. For both sources, filtering lowers the backlog a research office takes on at adoption. The recurring effort falls on researchers, and there OpenAlex costs them some \effResFlowGapHalf{} person-days a year more across the primary sample.

The effort also falls on different people. Researchers can reject a misattributed paper at a glance, but they receive decision requests as a series of notifications, each an interruption with a login attached, and because suggestions arrive a few at a time over a year the overhead of a session is paid for almost every decision. Batching notifications more coarsely does little when there is rarely more than one suggestion to batch, so the levers lie elsewhere: in filtering before anything reaches a researcher, in claiming automatically on an identifier that the researcher has asserted, and in keeping in the research office the investigations that a researcher's own knowledge does not shorten. Part of the burden on researchers is also not captured by counting minutes. A researcher who follows a notification has to leave the work in hand, and the login stands between them and the task they came to do. A step in that position is felt as an obstacle out of proportion to its length \citep{reynolds2020systemic}. People who are interrupted also tend to make up the lost time by working faster, at the price of greater stress \citep{mark2008cost}. A count of minutes per session therefore understates how a steady stream of notifications is experienced, even when each costs little time.

Even so, the workflow modelled here leaves researchers reviewing a source's guesses about their work and entering what it misses. There are many reasons to want researchers to take control of their own records: they are best placed to know what they wrote, and a record they maintain themselves is one for which responsibility is clear \citep{haak2012}. Taking control also removes most of the effort. A researcher who insists that their ORCID iD is recorded during the publishing process, so that the publisher deposits it with Crossref, and who connects their ORCID record to their CRIS profile, has new publications attributed without relying on a secondary source's disambiguation, and a CRIS such as Elements can claim such items automatically \citep{symplectic_autoclaim}. For that researcher there is nothing to review and nothing to enter, and the administrative burden on both the researcher and the research office falls to effectively zero. The effort model does not capture this route, because outputs attributed in this way never reach the feed as suggestions. How far it extends depends on how consistently researchers assert their identifiers and publishers carry them, which, as Section~\ref{sec:curationplace} discusses, remains uneven.

Two qualifications apply to these estimates. The backlog at adoption and the recurring effort of each further year are different quantities, one incurred once and the other every year, and should not simply be added together. And the single-feed assumption overstates entry in practice: for example, an Elements instance harvests several sources at once, a source that is weak on attribution can still serve as a supplementary feed, and secondary sources work alongside direct ingestion from the primary registries. Because the sample is weighted towards large institutions, figures for a whole research system are better obtained by scaling the per-article excess of unconfirmed attributions to its total output than by multiplying per-institution figures. Neither qualification alters the result, since OpenAlex requires more effort than Dimensions at the low, central and high settings alike, and even at the low setting the filtered backlog differs by some \effBacklogGapHalfLow{} person-days (\effOABacklogHalfLow{} against \effDimBacklogHalfLow{}). The integration effort is accordingly not a fixed property of a source, and it shrinks as ORCID, Crossref and DataCite become more complete, provided, of course, that a source makes use of all the primary information available to it.

\subsection{Where curation belongs}\label{sec:curationplace}

When national evaluations run on a secondary source, institutions may feel pressure to correct that source's copy of the record rather than strengthen the primary layer from which every source could draw. Because OpenAlex is published under a CC0 licence, effort spent correcting it is sometimes framed as a contribution to a public good. A correction applied to an algorithmic aggregation is, however, a patch on the output of a process that continues to run: reprocessing can overwrite it, and identifiers detached from the systems that generated them can come to point at different people as the algorithms beneath them are revised \citep{porter2024barcelona}, which is why corrections made at the primary source should always be preferred.

This suggests a property that an institution might reasonably ask of any secondary source: whether curation invested in the primary layer flows back into the source's own output. The Barcelona Declaration's commitments concerning institutional records such as those studied here invite the same question \citep{barcelona2024}. The two sources compared in this article differ in this respect. Dimensions associates its algorithmic author clusters with ORCID records through overlap of DOIs, so that a researcher's curated ORCID record flows back into its disambiguation \citep{hook2018}. OpenAlex's published pipeline anchors author identity on the ORCID iDs carried in the metadata of each work, chiefly in Crossref and DataCite records and on publishers' landing pages, filters them for reliability, and supplements them with user curation of ORCID iDs and of the links between authors and works. Importantly, an inspection of OpenAlex's source code shows that OpenAlex does not take assertions from ORCID directly \citep{ourresearch2026code}. Without the ability to benefit from the corrected ORCID record, corrections must be made directly into OpenAlex. OpenAlex's own reporting indicates the scale of community correction: a legacy author-correction form \citep{openalex_fix} collected 801 submissions over two years, replayed as some 10,300 curations \citep{openalex_members2026}, and requests to fix author profiles form its largest category of support, with a backlog of 584 actionable tickets worked through in mid-2026 \citep{openalex_cab2026}. The documented volume of curation requests is small by comparison with the \statOATailN{} identifiers beyond the conflation ceiling.

An institution's investment in ORCID strengthens every source that reads ORCID directly, and wider adoption of persistent identifiers is the most effective means available of reducing integration cost for all uses at once. Challenges for adoption remain, however, and uptake of ORCID varies between research systems and publishers \citep{porter2022orcid}. Further, although there is significant scope for wider adoption of ORCID, it is unreasonable to expect adoption ever to be complete. Some researchers withhold or restrict their records for reasons of privacy \citep{porterSR2025}, and in the European Union the processing of personal data in a CRIS is subject to the General Data Protection Regulation \citep{gdpr2016,schopfel2020ethics}. Where identifiers stop, attribution falls back on the institution's own curation, which is one reason to expect integration cost to remain material even in a well-identified research system.

\subsection{Identity errors as integrity risk}

The cost of identity errors to the wider research system needs to be considered separately. Integrity applications, such as screens for hyperprolific authorship \citep{ioannidis2018} or for fabricated collaboration networks \citep{porter2024networks}, consume author identity as their input, and their error rates are bounded below by that of the identity layer. Where profiles conflate, a genuine bad actor's signal can be diluted across identities until it falls below every threshold, while a researcher whose profile has absorbed another person's output can inherit a signal that was never theirs. A profile bearing the statistical signature of conflation is an artefact of an algorithm rather than evidence about the person behind it, so integrity tooling built on an identity layer with a known tendency to overmerge should require stronger evidence about a person than the identity layer alone can supply, and tool builders would do well to report which identity layer they build on, and its expected rate of confusion, as routinely as they report their own precision.

\subsection{Openness and the cost of infrastructure}

Suitability as an institutional record, the dimension measured here, is one of several considerations in choosing a source, alongside research security, research integrity and access to wider classes of data such as patents, none of which this paper measures; each, however, consumes the source's author identities. The results also bear on the framing of OpenAlex as open infrastructure, on which national bodies have begun to act: in Finland the education ministry, the Research Council of Finland and CSC have jointly piloted OpenAlex as a basis for national publication analyses, a move the ministry has endorsed \citep{okm2026}. The openness of a licence does not settle the cost of the infrastructure, and adopting any source for person-level use carries an integration cost that falls on institutions rather than on the body that chooses it. Durability of access is a further consideration. OpenAlex's coverage of recent affiliations has been reported to depend on harvesting publisher websites behind increasingly restrictive bot defences, latterly with commercial bot-evasion services, which makes it dependent both on the efficacy of that harvesting and on the community's willingness to accept the practice \citep{porter2026blog}; its full text comes from open-access copies or Internet Archive n-grams \citep{openalex_ngrams}; and its abstract coverage of closed-access output has fallen as publishers withdrew permission for harvested content \citep{kramer2024abstracts,tay2025}.  

Openness as infrastructure may be better judged by how effectively a source organises the open scholarly record, and how far it amplifies the community's work in correcting it, than by the licence on its data. On that reading proprietary systems are not excluded from contributing to open infrastructure, and the zone between freely available structured data and the commercially refined products built upon it, which \citet{hook2026market} calls the \emph{innovation annulus}, is a functional and lasting feature of the ecosystem. The percentages reported here are a snapshot of a disambiguation that is actively developed, and improvement in the specific figures would be welcome. The structural argument does not rest on them, although we would expect an openly licensed corpus disambiguated algorithmically at global scale, without an editorial correction loop, to face a harder version of the problem of merging and splitting than an index whose disambiguation is corrected through a loop with researcher-curated records, a hypothesis that a wider comparison could test. More broadly, researcher-level results resting on algorithmic identifiers are best validated against several sources, which the availability of sources in common environments such as Google BigQuery \citep{orcid2026bq,ds2025srad,orion2026}, and AI assistance in translating queries between schemas, should make routine.

%% file: sections/06_limitations.tex
\section{Limitations}\label{sec:limitations}

One of the limitations below marks an analysis that we regard as required before the findings are relied upon. First, the unconfirmed exposure of Section~\ref{sec:insttests} is read throughout as carrying false-positive risk, but its division between its two possible readings, genuine publications that the institution has not yet recorded and misattributions arising from poor author matches, has not been measured directly. A stratified manual audit of sampled unconfirmed attributions, by source, is the natural instrument, and the replication results of Section~\ref{sec:replication} should be weighted accordingly until it exists. The same audit would calibrate the effort model of Section~\ref{sec:effortmethod}, by measuring how many unconfirmed attributions need investigation and how long each takes.

Second, the overmerge comparison is presented descriptively, and its two comparisons run over different populations by construction (Section~\ref{sec:sourcetests}). The institutional tests address the main objection to this, since they measure both sources on the same researchers against a record that neither source produced, and they too find more error in OpenAlex: pooled across the primary sample it has the lower confirmed share (\statOAConfirmedPctPrimary\% against \statDimConfirmedPctPrimary\%) and the larger unconfirmed exposure (\statOANotInSystemPctPrimary\% against \statDimNotInSystemPctPrimary\%), and it retains fewer trusted profiles at a matched threshold at every one of the \statNPrimary{} primary institutions (Section~\ref{sec:trust}). A version of the anchored test restricted to authors identifiable under both anchor pairings, with a formal test of the asymmetry, would make the source-level comparison stand on its own, but the conclusions do not depend on it. The anchored populations are also conditioned in two ways that limit what the test can show. An ORCID record enters only with more than five DOIs in the common universe, so the benchmark says nothing about researchers early in their careers; and both the anchor and the measured profile must be linked to ORCID in their sources, so the anchored populations inherit ORCID's uneven adoption \citep{porter2022orcid}.

Third, each Discovery instance is a lower bound on its institution's true record. Institutional records are confirmed by people and therefore incomplete, so the ceiling on replication is the institutions' own completeness, which is why the first limitation matters. The records are also dated: the Discovery instances were last observed on \srcDiscoveryHarvest{}, so works published after that date cannot have been confirmed. They make up about \effOAUnconfLatePct\% of OpenAlex's unconfirmed attributions and \effDimUnconfLatePct\% of Dimensions', so they raise Dimensions' measured exposure more than OpenAlex's and do not favour Dimensions. The candidate pool from which the record is drawn also differs between institutions, since the licensed sources that an Elements instance harvests depend on its subscriptions. Where an institution licenses Dimensions as a harvest source, publications that Dimensions attributes correctly are more likely to have been proposed to the researcher and confirmed, which could raise its confirmed share. The effect extends beyond which candidates are proposed. Researchers can search Dimensions, Scopus and Web of Science by those sources' own author identifiers and can set such an identifier to claim matching items automatically \citep{symplectic_searchsettings,symplectic_autoclaim}, so that at an institution licensing Dimensions some confirmed links may rest on Dimensions' disambiguation rather than on a judgement about each item; OpenAlex is not an Elements data source and cannot benefit in the same way. Where links rest on a judgement about each item the effect bears on recall rather than precision, but automatic claiming on a Dimensions identifier would also let a Dimensions conflation enter the record unexamined and so raise Dimensions' measured precision. The restriction to Crossref journal articles limits the effect, since much of that universe is also within reach of the freely available sources; nonetheless, a comparison restricted to institutions that do not license Dimensions in Elements would bound the effect and should be reported. Such a comparison would require the informed participation of institutions that use Elements, since which sources an instance licenses, and how its researchers claim, is known only to the institution. Another approach would be to repeat the study on another CRIS, such as Elsevier's Pure, which draws on a different mix of sources. Automatic claiming on ORCID iDs similarly favours publications whose metadata carry them, which is neutral between the two sources but shapes the benchmark.

Fourth, the conflation ceilings of 40, and of 20 for the curation-grade check, are round-number judgements. The sensitivity analysis of Section~\ref{sec:sensitivity} indicates that the qualitative findings are robust to the choice, but the thresholds themselves are not derived quantities.

Fifth, institutions running Elements are concentrated in the United Kingdom and the wider anglophone research world, and the difficulty of disambiguating author names is known to vary with naming conventions that differ across languages and regions. The uniformity of the result should not be extrapolated to systems with materially different profiles of name commonality without measurement.

Sixth, the demonstration compares two sources chosen because they are built in the same way. Applying the same tests to other indexes built on open persistent-identifier infrastructure would test more widely the structural expectation of Section~\ref{sec:discussion}, that scale without an editorial correction loop predicts overmerging, and is the most valuable single extension of this work.

Finally, the effort model rests on analogues rather than on measurement. No published study measures how long it takes to accept or reject a suggested publication in a CRIS, so the unit times are borrowed from cataloguing, deposit and the screening of systematic reviews, which call for judgements of relevance rather than of authorship; researchers' decisions are charged the administrator's verification time, which is probably conservative for their own work; and two parameters, the time to investigate an attribution and the overhead of a researcher's session, are not measured at all and are varied across their ranges. Because the model counts only attribution decisions, and assumes that administrators keep their pace and accuracy through any number of them, its absolute figures are also a floor; the comparison between sources rests on their ratios. Effort is expressed in person-time rather than in money because curation practice and pay differ between institutions and regions, so that any single figure is an average over conditions that vary widely; the volume of records analysed makes that average a robust guide to the work involved, but not to its cost at any one institution. The recurring effort also rests on two choices of measurement. New works are timed by their first appearance in Dimensions, the only clock available for both sources, and an unconfirmed attribution is not necessarily wrong, so the probability that a new one will be confirmed is taken from the history of its identifier. Measuring these times directly, in each CRIS as it is used, would require collaboration with participating institutions, and should span more than one CRIS, such as Elements, Pure and Converis, so that the times do not reflect the interface of a single product.

%% file: sections/07_conclusion.tex
\section{Conclusion}\label{sec:conclusion}

Choosing a scientometric source for an institutional record commits the institution to a quantity of curation, and the method described here allows that quantity to be estimated in advance. An institution's own curated record, maintained for operational reasons and confirmed by the people it describes, serves as a benchmark against which any candidate source can be read; four tests, two of which need nothing but public data, characterise how a source lumps and splits identities and how much of the institution's record it can reproduce; and the result can be carried through to the person-days of research-office and researcher time that integrating the source would require, which for both sources are fewest when the feed is filtered before review.

Applied to OpenAlex and Dimensions across \statNPrimary{} institutions, the method finds a difference that is consistent in direction at every institution and material in size. The more lasting conclusion is that integration cost can be measured, depends on the use to which the data are put, and falls on institutions and researchers rather than on the bodies that choose the infrastructure. Curation invested in the primary registries and in institutions' own records serves every downstream use at once, within the limits that the protection of personal data places on identifiers, and it seems to us the most durable place for that effort to go.

%% file: sections/08_endmatter.tex
\section*{Acknowledgements}
The author thanks Daniel Hook for constructive internal peer review and for suggestions that improved the paper.

\section*{Author contributions}
Simon J. Porter: Conceptualization, Methodology, Software, Formal analysis, Data curation, Visualization, Writing -- original draft, Writing -- review \& editing.

\section*{Competing interests}
The author is an employee of Digital Science, which produces Dimensions and Altmetric and operates the Scientometric Researcher Access to Data programme \citep{ds2025srad}. Digital Science also owns Symplectic Elements, whose Discovery portals provide the institutional records used as ground truth in this study. Those records draw candidate publications from many sources, of which Dimensions is one that some institutions license; their person--publication links are confirmed by researchers or research offices, or claimed automatically on identifiers that researchers have asserted as their own. They draw on Dimensions' author identifiers only where a researcher has chosen to search or claim by them, which may favour Dimensions in the institutional tests (Section~\ref{sec:data} and Section~\ref{sec:limitations}); OpenAlex is not an Elements data source. OpenAlex is produced by the non-profit OurResearch, with which the author has no affiliation. To keep the comparison independent of these interests, the method and every reported quantity are reproducible from public data and published code, the tests can be run against any source, and the account of OpenAlex's pipeline rests on inspection of its public code rather than on privileged knowledge.

\section*{Funding information}
This work was funded by Digital Science as part of the author's employment.

\section*{Data availability}
The code, queries and validation checks used in this study are deposited on figshare \citep{porter2026code}. The institutional records themselves, although drawn from public Discovery portals, are not redistributed, since their republication is properly a decision for each institution. Instead, the deposit includes a harvester that retrieves the same fields from the public Discovery portals, with the list of portals it was run on, so that the records can be retrieved afresh and the study population reconstructed.

%% file: main.bbl
\begin{thebibliography}{70}
\providecommand{\natexlab}[1]{#1}
\providecommand{\url}[1]{\texttt{#1}}
\expandafter\ifx\csname urlstyle\endcsname\relax
  \providecommand{\doi}[1]{doi: #1}\else
  \providecommand{\doi}{doi: \begingroup \urlstyle{rm}\Url}\fi

\bibitem[Alonso-{\'A}lvarez and van Eck(2025)]{alonso2025}
P.~Alonso-{\'A}lvarez and N.~J. van Eck.
\newblock Coverage and metadata completeness and accuracy of {African} research
  publications in {OpenAlex}: A comparative analysis.
\newblock \emph{Quantitative Science Studies}, 6:\penalty0 1336--1357, 2025.
\newblock \doi{10.1162/QSS.a.396}.

\bibitem[Azeroual and Sch{\"o}pfel(2019)]{azeroual2019}
O.~Azeroual and J.~Sch{\"o}pfel.
\newblock Quality issues of {CRIS} data: An exploratory investigation with
  universities from twelve countries.
\newblock \emph{Publications}, 7\penalty0 (1):\penalty0 14, 2019.
\newblock \doi{10.3390/publications7010014}.

\bibitem[Azeroual et~al.(2018)Azeroual, Saake, and Schallehn]{azeroual2018}
O.~Azeroual, G.~Saake, and E.~Schallehn.
\newblock Analyzing data quality issues in research information systems via
  data profiling.
\newblock \emph{International Journal of Information Management}, 41:\penalty0
  50--56, 2018.
\newblock \doi{10.1016/j.ijinfomgt.2018.02.007}.

\bibitem[{Barcelona Declaration on Open Research
  Information}(2024)]{barcelona2024}
{Barcelona Declaration on Open Research Information}.
\newblock Barcelona declaration on open research information.
\newblock Version 1.0, 16 April 2024, Zenodo, 2024.
\newblock Edited by B. Kramer, C. Neylon and L. Waltman.

\bibitem[Brase(2009)]{brase2009}
J.~Brase.
\newblock {DataCite} -- a global registration agency for research data.
\newblock In \emph{2009 Fourth International Conference on Cooperation and
  Promotion of Information Resources in Science and Technology (COINFO '09)},
  pages 257--261, Beijing, 2009. IEEE.
\newblock \doi{10.1109/COINFO.2009.66}.

\bibitem[Brown et~al.(2022)Brown, Jones, Meadows, and Murphy]{brown2022}
J.~Brown, P.~Jones, A.~Meadows, and F.~Murphy.
\newblock Incentives to invest in identifiers: A cost-benefit analysis of
  persistent identifiers in {Australian} research systems.
\newblock Technical report, MoreBrains Cooperative, for the Australian Research
  Data Commons and the Australian Access Federation, 2022.

\bibitem[Bryant et~al.(2018)Bryant, Clements, de~Castro, Cantrell, Dortmund,
  Fransen, Gallagher, and Mennielli]{bryant2018}
R.~Bryant, A.~Clements, P.~de~Castro, J.~Cantrell, A.~Dortmund, J.~Fransen,
  P.~Gallagher, and M.~Mennielli.
\newblock Practices and patterns in research information management: Findings
  from a global survey.
\newblock Technical report, OCLC Research, Dublin, OH, 2018.

\bibitem[Caron and van Eck(2014)]{caron2014}
E.~Caron and N.~J. van Eck.
\newblock Large scale author name disambiguation using rule-based scoring and
  clustering.
\newblock In Ed~Noyons, editor, \emph{Proceedings of the Science and Technology
  Indicators Conference 2014 (STI 2014): Context counts: Pathways to master big
  and little data}, pages 79--86, Leiden, 2014. Universiteit Leiden.

\bibitem[{COPE} and {STM}(2022)]{cope2022}
{COPE} and {STM}.
\newblock Paper mills: Research report from {COPE} \& {STM}.
\newblock Technical report, Committee on Publication Ethics and STM, 2022.

\bibitem[Culbert et~al.(2025)Culbert, Hobert, Jahn, Haupka, Schmidt, Donner,
  and Mayr]{culbert2025}
J.~H. Culbert, A.~Hobert, N.~Jahn, N.~Haupka, M.~Schmidt, P.~Donner, and
  P.~Mayr.
\newblock Reference coverage analysis of {OpenAlex} compared to {Web of
  Science} and {Scopus}.
\newblock \emph{Scientometrics}, 130\penalty0 (4):\penalty0 2475--2492, 2025.
\newblock \doi{10.1007/s11192-025-05293-3}.

\bibitem[D'Angelo and van Eck(2020)]{dangelo2020}
C.~A. D'Angelo and N.~J. van Eck.
\newblock Collecting large-scale publication data at the level of individual
  researchers: a practical proposal for author name disambiguation.
\newblock \emph{Scientometrics}, 123\penalty0 (2):\penalty0 883--907, 2020.
\newblock \doi{10.1007/s11192-020-03410-y}.

\bibitem[{Digital Science}(2025)]{ds2025srad}
{Digital Science}.
\newblock Digital science relaunches {Scientometric Researcher Access to Data
  (SRAD)} program.
\newblock Digital Science press release, 22 July 2025, 2025.
\newblock URL
  \url{https://www.digital-science.com/press-releases/digital-science-relaunches-scientometric-researcher-access-to-data-srad-program/}.
\newblock Accessed 28 October 2026.

\bibitem[{Digital Science}(2026)]{dimensions_bq_docs}
{Digital Science}.
\newblock Dimensions on {Google BigQuery}: documentation.
\newblock Digital Science, 2026.
\newblock URL \url{https://docs.dimensions.ai/bigquery/}.
\newblock Publications and researchers tables; the publications field
  date\_inserted records the ``date when the record was inserted into
  Dimensions''. Accessed 28 October 2026.

\bibitem[Else and Van~Noorden(2021)]{else2021}
H.~Else and R.~Van~Noorden.
\newblock The fight against fake-paper factories that churn out sham science.
\newblock \emph{Nature}, 591\penalty0 (7851):\penalty0 516--519, 2021.
\newblock \doi{10.1038/d41586-021-00733-5}.

\bibitem[{European Parliament and Council of the European
  Union}(2016)]{gdpr2016}
{European Parliament and Council of the European Union}.
\newblock Regulation ({EU}) 2016/679 of the {European Parliament} and of the
  {Council} of 27 {April} 2016 on the protection of natural persons with regard
  to the processing of personal data and on the free movement of such data, and
  repealing {Directive} 95/46/{EC} ({General Data Protection Regulation}).
\newblock Official Journal of the European Union, L 119, 4 May 2016, pp. 1--88,
  2016.
\newblock URL \url{https://eur-lex.europa.eu/eli/reg/2016/679/oj/eng}.

\bibitem[{FCCN -- FCT Digital Services}(2024)]{fccn2024cienciavitae}
{FCCN -- FCT Digital Services}.
\newblock Nova funcionalidade do {CI\^ENCIAVITAE} alivia carga burocr\'atica
  sobre os investigadores.
\newblock FCCN news, 4 September 2024, 2024.
\newblock URL
  \url{https://www.fccn.pt/atualidade/funcionalidade-cienciavitae-alivia-carga-burocratica/}.
\newblock Accessed 25 October 2026.

\bibitem[Ferreira et~al.(2012)Ferreira, Gon{\c{c}}alves, and
  Laender]{ferreira2012}
A.~A. Ferreira, M.~A. Gon{\c{c}}alves, and A.~H.~F. Laender.
\newblock A brief survey of automatic methods for author name disambiguation.
\newblock \emph{ACM SIGMOD Record}, 41\penalty0 (2):\penalty0 15--26, 2012.
\newblock \doi{10.1145/2350036.2350040}.

\bibitem[Fister et~al.(2016)Fister, Fister, and Perc]{fister2016}
I.~Fister, Jr., I.~Fister, and M.~Perc.
\newblock Toward the discovery of citation cartels in citation networks.
\newblock \emph{Frontiers in Physics}, 4:\penalty0 49, 2016.
\newblock \doi{10.3389/fphy.2016.00049}.

\bibitem[Gurney(2022)]{gurney2022}
T.~Gurney.
\newblock The changing scope of data quality and fit for purpose: evolution and
  adaption of a {CRIS} solution.
\newblock \emph{Procedia Computer Science}, 211:\penalty0 201--205, 2022.
\newblock \doi{10.1016/j.procs.2022.10.192}.

\bibitem[Haak et~al.(2012)Haak, Fenner, Paglione, Pentz, and Ratner]{haak2012}
L.~L. Haak, M.~Fenner, L.~Paglione, E.~Pentz, and H.~Ratner.
\newblock {ORCID}: a system to uniquely identify researchers.
\newblock \emph{Learned Publishing}, 25\penalty0 (4):\penalty0 259--264, 2012.
\newblock \doi{10.1087/20120404}.

\bibitem[Hendricks et~al.(2020)Hendricks, Tkaczyk, Lin, and
  Feeney]{hendricks2020}
G.~Hendricks, D.~Tkaczyk, J.~Lin, and P.~Feeney.
\newblock Crossref: The sustainable source of community-owned scholarly
  metadata.
\newblock \emph{Quantitative Science Studies}, 1\penalty0 (1):\penalty0
  414--427, 2020.
\newblock \doi{10.1162/qss_a_00022}.

\bibitem[Herzog et~al.(2020)Herzog, Hook, and Konkiel]{herzog2020}
C.~Herzog, D.~Hook, and S.~Konkiel.
\newblock Dimensions: Bringing down barriers between scientometricians and
  data.
\newblock \emph{Quantitative Science Studies}, 1\penalty0 (1):\penalty0
  387--395, 2020.
\newblock \doi{10.1162/qss_a_00020}.

\bibitem[Hook(2026)]{hook2026market}
D.~W. Hook.
\newblock Market dynamics, governance and open research metadata in the {AI}
  era.
\newblock arXiv preprint arXiv:2604.19507v2, 2026.
\newblock URL \url{https://arxiv.org/abs/2604.19507}.

\bibitem[Hook et~al.(2018)Hook, Porter, and Herzog]{hook2018}
D.~W. Hook, S.~J. Porter, and C.~Herzog.
\newblock Dimensions: Building context for search and evaluation.
\newblock \emph{Frontiers in Research Metrics and Analytics}, 3:\penalty0 23,
  2018.
\newblock \doi{10.3389/frma.2018.00023}.

\bibitem[Hook et~al.(2021)Hook, Porter, Draux, and Herzog]{hook2021realtime}
Daniel~W. Hook, Simon~J. Porter, H{\'e}l{\`e}ne Draux, and Christian~T. Herzog.
\newblock Real-time bibliometrics: {Dimensions} as a resource for analyzing
  aspects of {COVID}-19.
\newblock \emph{Frontiers in Research Metrics and Analytics}, 5:\penalty0
  595299, 2021.
\newblock \doi{10.3389/frma.2020.595299}.

\bibitem[Howard et~al.(2020)Howard, Phillips, Tandon, Maharana, Elmore, Mav,
  Sedykh, Thayer, Merrick, Walker, Rooney, and Shah]{howard2020swift}
Brian~E. Howard, Jason Phillips, Arpit Tandon, Adyasha Maharana, Rebecca
  Elmore, Deepak Mav, Alex Sedykh, Kristina Thayer, B.~Alex Merrick, Vickie
  Walker, Andrew Rooney, and Ruchir~R. Shah.
\newblock {SWIFT-Active Screener}: Accelerated document screening through
  active learning and integrated recall estimation.
\newblock \emph{Environment International}, 138:\penalty0 105623, 2020.
\newblock \doi{10.1016/j.envint.2020.105623}.

\bibitem[Ioannidis et~al.(2018)Ioannidis, Klavans, and Boyack]{ioannidis2018}
J.~P.~A. Ioannidis, R.~Klavans, and K.~W. Boyack.
\newblock Thousands of scientists publish a paper every five days.
\newblock \emph{Nature}, 561\penalty0 (7722):\penalty0 167--169, 2018.
\newblock \doi{10.1038/d41586-018-06185-8}.

\bibitem[Iqbal and Horvitz(2007)]{iqbal2007disruption}
Shamsi~T. Iqbal and Eric Horvitz.
\newblock Disruption and recovery of computing tasks: Field study, analysis,
  and directions.
\newblock In \emph{Proceedings of the SIGCHI Conference on Human Factors in
  Computing Systems (CHI '07)}, pages 677--686. ACM, 2007.
\newblock \doi{10.1145/1240624.1240730}.

\bibitem[Kramer(2024)]{kramer2024abstracts}
B.~Kramer.
\newblock More open abstracts? comparing abstract coverage in {Crossref} and
  {OpenAlex}.
\newblock Sesame Open Science, 24 November 2024, 2024.
\newblock URL
  \url{https://bmkramer.github.io/SesameOpenScience_site/thought/202411_open_abstracts/}.
\newblock Accessed 22 July 2026.

\bibitem[Krupinski et~al.(2010)Krupinski, Berbaum, Caldwell, Schartz, and
  Kim]{krupinski2010long}
Elizabeth~A. Krupinski, Kevin~S. Berbaum, Robert~T. Caldwell, Kevin~M. Schartz,
  and John Kim.
\newblock Long radiology workdays reduce detection and accommodation accuracy.
\newblock \emph{Journal of the American College of Radiology}, 7\penalty0
  (9):\penalty0 698--704, 2010.
\newblock \doi{10.1016/j.jacr.2010.03.004}.

\bibitem[Linder et~al.(2014)Linder, Doctor, Friedberg, Reyes~Nieva, Birks,
  Meeker, and Fox]{linder2014time}
Jeffrey~A. Linder, Jason~N. Doctor, Mark~W. Friedberg, Harry Reyes~Nieva,
  Caroline Birks, Daniella Meeker, and Craig~R. Fox.
\newblock Time of day and the decision to prescribe antibiotics.
\newblock \emph{JAMA Internal Medicine}, 174\penalty0 (12):\penalty0
  2029--2031, 2014.
\newblock \doi{10.1001/jamainternmed.2014.5225}.

\bibitem[Mark et~al.(2005)Mark, Gonzalez, and Harris]{mark2005notask}
Gloria Mark, Victor~M. Gonzalez, and Justin Harris.
\newblock No task left behind? examining the nature of fragmented work.
\newblock In \emph{Proceedings of the SIGCHI Conference on Human Factors in
  Computing Systems (CHI '05)}, pages 321--330. ACM, 2005.
\newblock \doi{10.1145/1054972.1055017}.

\bibitem[Mark et~al.(2008)Mark, Gudith, and Klocke]{mark2008cost}
Gloria Mark, Daniela Gudith, and Ulrich Klocke.
\newblock The cost of interrupted work: More speed and stress.
\newblock In \emph{Proceedings of the SIGCHI Conference on Human Factors in
  Computing Systems (CHI '08)}, pages 107--110. ACM, 2008.
\newblock \doi{10.1145/1357054.1357072}.

\bibitem[{Ministry of Education and Culture, Finland}(2026)]{okm2026}
{Ministry of Education and Culture, Finland}.
\newblock Ministry of education and culture endorses national move towards open
  source research data.
\newblock News item, 30 April 2026, 2026.
\newblock URL
  \url{https://okm.fi/en/-/ministry-of-education-and-culture-endorses-national-move-towards-open-source-research-data}.
\newblock Accessed 21 July 2026.

\bibitem[O'Blenis(2017)]{oblenis2017vote}
Peter O'Blenis.
\newblock Vote or classify: Which is faster for title and abstract screening?
\newblock DistillerSR blog, 22 June 2017, 2017.
\newblock URL
  \url{https://blog.distillersr.com/resources/blog/vote-or-classify-which-is-faster-for-title-and-abstract-screening}.
\newblock Accessed 25 October 2026.

\bibitem[{OpenAlex}(2026{\natexlab{a}})]{openalex_cab2026}
{OpenAlex}.
\newblock {OpenAlex} community advisory board -- {July} 2026 (retrospective).
\newblock OpenAlex town hall notes, 9 July 2026, 2026{\natexlab{a}}.
\newblock URL
  \url{https://github.com/ourresearch/town-hall-notes/blob/main/2026-07-09-cab/retro.md}.
\newblock Accessed 24 July 2026.

\bibitem[{OpenAlex}(2026{\natexlab{b}})]{openalex_members2026}
{OpenAlex}.
\newblock {OpenAlex} members meeting -- {May} 2026 (retrospective).
\newblock OpenAlex town hall notes, 27 May 2026, 2026{\natexlab{b}}.
\newblock URL
  \url{https://github.com/ourresearch/town-hall-notes/blob/main/2026-05-27-members-meeting/retro.md}.
\newblock Accessed 24 July 2026.

\bibitem[{OpenAlex}(n.d.{\natexlab{a}})]{openalex_fix}
{OpenAlex}.
\newblock How can {I} fix errors in an {OpenAlex} author profile?
\newblock OpenAlex help centre (since replaced by ``Fixing authors''),
  n.d.{\natexlab{a}}.
\newblock URL
  \url{https://web.archive.org/web/20260218035317/https://help.openalex.org/hc/en-us/articles/27283405287319-How-can-I-fix-errors-in-an-OpenAlex-author-profile}.
\newblock Archived 18 February 2026.

\bibitem[{OpenAlex}(n.d.{\natexlab{b}})]{openalex_ngrams}
{OpenAlex}.
\newblock Get {N}-grams.
\newblock OpenAlex technical documentation (archived repository
  ourresearch/openalex-docs), n.d.{\natexlab{b}}.
\newblock URL
  \url{https://github.com/ourresearch/openalex-docs/blob/main/api-entities/works/get-n-grams.md}.
\newblock Repository since archived; accessed 22 July 2026.

\bibitem[{ORCID}(2026)]{orcid2026bq}
{ORCID}.
\newblock {ORCID} public data file on {Google BigQuery}.
\newblock ORCID, 2026.
\newblock URL
  \url{https://info.orcid.org/orcid-public-data-file-on-google-big-query/}.
\newblock Describes the ORCID public data files hosted on Google BigQuery; this
  study uses the 2026 summaries file. Accessed 25 October 2026.

\bibitem[{ORION-DBs}(2026)]{orion2026}
{ORION-DBs}.
\newblock {ORION-DBs}: Open research information on {BigQuery}.
\newblock Community registry of open scholarly metadata datasets on Google
  BigQuery, 2026.
\newblock URL \url{https://orion-dbs.community/}.
\newblock OpenAlex snapshot subugoe-collaborative.openalex\_walden, hosted by
  SUB G{\"o}ttingen, July 2026 load; accessed 25 July 2026.

\bibitem[{OurResearch}(2026)]{ourresearch2026code}
{OurResearch}.
\newblock {OpenAlex} data pipeline source code (repositories openalex-ingest
  and openalex-walden).
\newblock GitHub, 2026.
\newblock URL
  \url{https://github.com/ourresearch/openalex-walden/tree/d60a2a70f6504c5ac5090b103ab9bfb9beb9d249}.
\newblock Static inspection of the main branches as of 25 October 2026:
  openalex-walden at commit d60a2a7 (URL above) and openalex-ingest at commit
  f6ee9ad,
  \url{https://github.com/ourresearch/openalex-ingest/tree/f6ee9adec7b9e2dec450c82c945a812f7265c302}.

\bibitem[{Overton}(2024)]{overton2026}
{Overton}.
\newblock {OpenAlex} and researcher profiles.
\newblock Overton knowledge base, 22 May 2024, updated 12 August 2026, 2024.
\newblock URL
  \url{https://help.overton.io/article/open-alex-and-researcher-profiles/}.
\newblock Accessed 28 October 2026.

\bibitem[Porter(2022)]{porter2022orcid}
S.~J. Porter.
\newblock Measuring research information citizenship across {ORCID} practice.
\newblock \emph{Frontiers in Research Metrics and Analytics}, 7:\penalty0
  779097, 2022.
\newblock \doi{10.3389/frma.2022.779097}.

\bibitem[Porter(2024)]{porter2024barcelona}
S.~J. Porter.
\newblock The {Barcelona Declaration}\ldots exploring our responsibilities as
  metadata consumers.
\newblock Digital Science blog, 10 July 2024, 2024.
\newblock URL
  \url{https://www.digital-science.com/blog/the-barcelona-declaration-exploring-our-responsibilities-as-metadata-consumers/}.
\newblock Accessed 21 July 2026.

\bibitem[Porter(2026{\natexlab{a}})]{porter2026blog}
S.~J. Porter.
\newblock No shortcuts to research information citizenship.
\newblock Digital Science blog, 11 February 2026, 2026{\natexlab{a}}.
\newblock URL
  \url{https://www.digital-science.com/blog/no-shortcuts-to-research-information-citizenship/}.
\newblock Accessed 21 July 2026.

\bibitem[Porter and McIntosh(2024)]{porter2024networks}
S.~J. Porter and L.~D. McIntosh.
\newblock Identifying fabricated networks within authorship-for-sale
  enterprises.
\newblock \emph{Scientific Reports}, 14:\penalty0 29569, 2024.
\newblock \doi{10.1038/s41598-024-71230-8}.

\bibitem[Porter et~al.(2025)Porter, Umbach, and Willis]{porterSR2025}
S.~R. Porter, P.~D. Umbach, and C.~Willis.
\newblock Understanding {ORCID} adoption among academic researchers.
\newblock \emph{Scientometrics}, 130\penalty0 (5):\penalty0 2783--2797, 2025.
\newblock \doi{10.1007/s11192-025-05300-7}.

\bibitem[Porter(2026{\natexlab{b}})]{porter2026code}
Simon~J. Porter.
\newblock Code for ``{What It Costs to Trust a Source: A Reproducible Method
  for Institutional Assessment of Author Disambiguation}''.
\newblock figshare, 2026{\natexlab{b}}.
\newblock URL \url{https://doi.org/10.6084/m9.figshare.34831479}.
\newblock Analysis code, queries and validation checks, with a harvester for
  the Discovery records; MIT licence.

\bibitem[Priem et~al.(2022)Priem, Piwowar, and Orr]{priem2022}
J.~Priem, H.~Piwowar, and R.~Orr.
\newblock {OpenAlex}: A fully-open index of scholarly works, authors, venues,
  institutions, and concepts.
\newblock arXiv preprint arXiv:2205.01833; submitted to the 26th International
  Conference on Science, Technology and Innovation Indicators (STI 2022), 2022.
\newblock URL \url{https://arxiv.org/abs/2205.01833}.

\bibitem[Reynolds et~al.(2020)Reynolds, Samarin, Barnes, Judd, Mason, Bailey,
  and Egelman]{reynolds2020systemic}
Joshua Reynolds, Nikita Samarin, Joseph Barnes, Taylor Judd, Joshua Mason,
  Michael Bailey, and Serge Egelman.
\newblock Empirical measurement of systemic {2FA} usability.
\newblock In \emph{Proceedings of the 29th {USENIX} Security Symposium}, pages
  127--142. {USENIX} Association, 2020.
\newblock URL
  \url{https://www.usenix.org/conference/usenixsecurity20/presentation/reynolds}.

\bibitem[Scherer et~al.(2020)Scherer, Byrne, Hahnel, and Valen]{scherer2020}
D.~Scherer, K.~Byrne, M.~Hahnel, and D.~Valen.
\newblock Collaborative approaches to integrate repositories within the
  research information ecosystem: Creating bridges for common goals.
\newblock \emph{The Serials Librarian}, 78\penalty0 (1--4):\penalty0 181--190,
  2020.
\newblock \doi{10.1080/0361526X.2020.1728169}.

\bibitem[Sch{\"o}pfel et~al.(2017)Sch{\"o}pfel, Prost, and
  Rebouillat]{schopfel2017}
J.~Sch{\"o}pfel, H.~Prost, and V.~Rebouillat.
\newblock Research data in current research information systems.
\newblock \emph{Procedia Computer Science}, 106:\penalty0 305--320, 2017.
\newblock \doi{10.1016/j.procs.2017.03.030}.

\bibitem[Sch{\"o}pfel et~al.(2020)Sch{\"o}pfel, Azeroual, and
  Jungbauer-Gans]{schopfel2020ethics}
J.~Sch{\"o}pfel, O.~Azeroual, and M.~Jungbauer-Gans.
\newblock Research ethics, open science and {CRIS}.
\newblock \emph{Publications}, 8\penalty0 (4):\penalty0 51, 2020.
\newblock \doi{10.3390/publications8040051}.

\bibitem[Siguenza-Guzman et~al.(2014)Siguenza-Guzman, Van~den Abbeele, and
  Cattrysse]{siguenzaguzman2014cataloguing}
Lorena Siguenza-Guzman, Alexandra Van~den Abbeele, and Dirk Cattrysse.
\newblock Improving library management by using cost analysis tools: A case
  study for cataloguing processes.
\newblock \emph{LIBER Quarterly}, 23\penalty0 (3):\penalty0 160--186, 2014.
\newblock \doi{10.18352/lq.8558}.

\bibitem[Sinha et~al.(2015)Sinha, Shen, Song, Ma, Eide, Hsu, and
  Wang]{sinha2015}
A.~Sinha, Z.~Shen, Y.~Song, H.~Ma, D.~Eide, B.-J. Hsu, and K.~Wang.
\newblock An overview of {Microsoft Academic Service} ({MAS}) and applications.
\newblock In \emph{Proceedings of the 24th International Conference on World
  Wide Web (WWW '15 Companion)}, pages 243--246, Florence, 2015. ACM.
\newblock \doi{10.1145/2740908.2742839}.

\bibitem[Smalheiser and Torvik(2009)]{smalheiser2009}
N.~R. Smalheiser and V.~I. Torvik.
\newblock Author name disambiguation.
\newblock \emph{Annual Review of Information Science and Technology},
  43\penalty0 (1):\penalty0 1--43, 2009.
\newblock \doi{10.1002/aris.2009.1440430113}.

\bibitem[{Symplectic}(2016)]{symplectic2016qmul}
{Symplectic}.
\newblock Case study: Open access workflow improvements at {QMUL}.
\newblock Symplectic, 9 March 2016, reposted from the opeNWorks blog of 2 March
  2016, 2016.
\newblock URL
  \url{https://www.symplectic.co.uk/blog/case-study-open-access-workflow-improvements-at-qmul/}.
\newblock Accessed 25 October 2026.

\bibitem[{Symplectic}(2025)]{symplectic_searchsettings}
{Symplectic}.
\newblock Troubleshooting guide for search settings.
\newblock Symplectic Support Centre, edited 27 June 2025, 2025.
\newblock URL \url{https://support.symplectic.co.uk/en/articles/11297537}.
\newblock Accessed 25 October 2026.

\bibitem[{Symplectic}(2026{\natexlab{a}})]{symplectic_autoclaim}
{Symplectic}.
\newblock Automatic claiming in elements.
\newblock Symplectic Support Centre, edited 20 August 2026, 2026{\natexlab{a}}.
\newblock URL \url{https://support.symplectic.co.uk/en/articles/11299457}.
\newblock Accessed 25 October 2026.

\bibitem[{Symplectic}(2026{\natexlab{b}})]{symplectic_preferred}
{Symplectic}.
\newblock How does preferred record functionality work?
\newblock Symplectic Support Centre, edited 27 August 2026, 2026{\natexlab{b}}.
\newblock URL \url{https://support.symplectic.co.uk/en/articles/12051649}.
\newblock Accessed 25 October 2026.

\bibitem[{Symplectic}(n.d.)]{symplectic_datasources}
{Symplectic}.
\newblock Elements data sources and configuration.
\newblock Symplectic Support Centre, n.d.
\newblock URL \url{https://support.symplectic.co.uk/en/articles/11292225}.
\newblock Accessed 25 October 2026.

\bibitem[Tay(2025)]{tay2025}
A.~Tay.
\newblock The petrol tank for {AI} discovery might be running dry as publishers
  close access to scholarly content such as abstracts due to {AI} incentives.
\newblock Aaron Tay's Musings about Librarianship (Substack), 27 September
  2025, 2025.
\newblock URL
  \url{https://aarontay.substack.com/p/the-petrol-tank-for-ai-discovery}.
\newblock Accessed 22 July 2026.

\bibitem[Tekles and Bornmann(2020)]{tekles2020}
A.~Tekles and L.~Bornmann.
\newblock Author name disambiguation of bibliometric data: A comparison of
  several unsupervised approaches.
\newblock \emph{Quantitative Science Studies}, 1\penalty0 (4):\penalty0
  1510--1528, 2020.
\newblock \doi{10.1162/qss_a_00081}.

\bibitem[Trad et~al.(2025)Trad, Yammine, Charafeddine, Chakhtoura, Rahme,
  El-Hajj~Fuleihan, and Chehab]{trad2025streamlining}
Fouad Trad, Ryan Yammine, Jana Charafeddine, Marlene Chakhtoura, Maya Rahme,
  Ghada El-Hajj~Fuleihan, and Ali Chehab.
\newblock Streamlining systematic reviews with large language models using
  prompt engineering and retrieval augmented generation.
\newblock \emph{BMC Medical Research Methodology}, 25\penalty0 (1):\penalty0
  130, 2025.
\newblock \doi{10.1186/s12874-025-02583-5}.

\bibitem[{University of Alabama Libraries}(2026)]{ua_ir_deposit}
{University of Alabama Libraries}.
\newblock Deposit guidelines, institutional repository guide.
\newblock Research Guides at University of Alabama, last updated 14 April 2026,
  2026.
\newblock URL \url{https://guides.lib.ua.edu/c.php?g=710604&p=9893959}.
\newblock Accessed 25 October 2026.

\bibitem[Visser et~al.(2021)Visser, van Eck, and Waltman]{visser2021}
M.~Visser, N.~J. van Eck, and L.~Waltman.
\newblock Large-scale comparison of bibliographic data sources: {Scopus}, {Web
  of Science}, {Dimensions}, {Crossref}, and {Microsoft Academic}.
\newblock \emph{Quantitative Science Studies}, 2\penalty0 (1):\penalty0 20--41,
  2021.
\newblock \doi{10.1162/qss_a_00112}.

\bibitem[Wang et~al.(2020)Wang, Shen, Huang, Wu, Dong, and Kanakia]{wang2020}
K.~Wang, Z.~Shen, C.~Huang, C.-H. Wu, Y.~Dong, and A.~Kanakia.
\newblock {Microsoft Academic Graph}: When experts are not enough.
\newblock \emph{Quantitative Science Studies}, 1\penalty0 (1):\penalty0
  396--413, 2020.
\newblock \doi{10.1162/qss_a_00021}.

\bibitem[Wilkinson et~al.(2016)Wilkinson, Dumontier, Aalbersberg,
  et~al.]{wilkinson2016}
M.~D. Wilkinson, M.~Dumontier, I.~J. Aalbersberg, et~al.
\newblock The {FAIR} guiding principles for scientific data management and
  stewardship.
\newblock \emph{Scientific Data}, 3\penalty0 (1):\penalty0 160018, 2016.
\newblock \doi{10.1038/sdata.2016.18}.

\bibitem[Zhao and Chen(2025)]{zhao2025}
R.~Zhao and Y.~Chen.
\newblock Accuracy assessment of {OpenAlex} and {Clarivate} {Scholar ID} with
  an {LLM}-assisted benchmark.
\newblock arXiv preprint arXiv:2502.11610, 2025.

\end{thebibliography}
